\documentstyle[12pt]{article}
\input psfig.sty
\def\nn{\noindent}
\def\Re{{\cal R \mskip-4mu \lower.1ex \hbox{\it e}\,}}
\def\Im{{\cal I \mskip-5mu \lower.1ex \hbox{\it m}\,}}
\def\ie{{\it i.e.}}
\def\eg{{\it e.g.}}

\def\etal{{\it et al.}}

\def\sub#1{_{\lower.25ex\hbox{$\scriptstyle#1$}}}
\def\tev{\,{\ifmmode\mathrm {TeV}\else TeV\fi}}
\def\gev{\,{\ifmmode\mathrm {GeV}\else GeV\fi}}
\def\mev{\,{\ifmmode\mathrm {MeV}\else MeV\fi}}
\def\to{\rightarrow}

\def\subw{_{\rm w}}
\def\mh{\ifmmode m\sbl H \else $m\sbl H$\fi}
\def\mch{\ifmmode m_{H^\pm} \else $m_{H^\pm}$\fi}
\def\mt{\ifmmode m_t\else $m_t$\fi}
\def\mc{\ifmmode m_c\else $m_c$\fi}
\def\mz{\ifmmode M_Z\else $M_Z$\fi}
\def\mw{\ifmmode M_W\else $M_W$\fi}
\def\mws{\ifmmode M_W^2 \else $M_W^2$\fi}
\def\mhs{\ifmmode m_H^2 \else $m_H^2$\fi}   
\def\mzs{\ifmmode M_Z^2 \else $M_Z^2$\fi}
\def\mts{\ifmmode m_t^2 \else $m_t^2$\fi}
\def\mcs{\ifmmode m_c^2 \else $m_c^2$\fi}
\def\mchs{\ifmmode m_{H^\pm}^2 \else $m_{H^\pm}^2$\fi}
\def\ztwo{\ifmmode Z_2\else $Z_2$\fi}
\def\zone{\ifmmode Z_1\else $Z_1$\fi}
\def\mtwo{\ifmmode M_2\else $M_2$\fi}
\def\mone{\ifmmode M_1\else $M_1$\fi}
\def\tb{\ifmmode \tan\beta \else $\tan\beta$\fi}
\def\xw{\ifmmode x\subw\else $x\subw$\fi}
\def\ch{\ifmmode H^\pm \else $H^\pm$\fi}
\def\lum{\ifmmode {\cal L}\else ${\cal L}$\fi}
\def\inpb{\,{\ifmmode {\mathrm {pb}}^{-1}\else ${\mathrm {pb}}^{-1}$\fi}}
\def\infb{\,{\ifmmode {\mathrm {fb}}^{-1}\else ${\mathrm {fb}}^{-1}$\fi}}
\def\epem{\ifmmode e^+e^-\else $e^+e^-$\fi}
\def\ppb{\ifmmode \bar pp\else $\bar pp$\fi}
\def\bsg{\ifmmode B\to X_s\gamma\else $B\to X_s\gamma$\fi}
\def\bsll{\ifmmode B\to X_s\ell^+\ell^-\else $B\to X_s\ell^+\ell^-$\fi}
\def\bstt{\ifmmode B\to X_s\tau^+\tau^-\else $B\to X_s\tau^+\tau^-$\fi}
\def\lamt{\ifmmode \tilde\lambda\else $\tilde\lambda$\fi}
\def\shat{\ifmmode \hat s\else $\hat s$\fi}
\def\that{\ifmmode \hat t\else $\hat t$\fi}
\def\uhat{\ifmmode \hat u\else $\hat u$\fi}

\newskip\zatskip \zatskip=0pt plus0pt minus0pt
\def\matth{\mathsurround=0pt}

\def\atversim#1#2{\lower0.7ex\vbox{\baselineskip\zatskip\lineskip\zatskip
  \lineskiplimit 0pt\ialign{$\matth#1\hfil##\hfil$\crcr#2\crcr\sim\crcr}}}

\renewcommand{\thefootnote}{\fnsymbol{footnote}}

\begin{document} \begin{titlepage} 
\rightline{\vbox{\halign{&#\hfil\cr
&SLAC-PUB-7838\cr
&June 1998\cr}}}
\begin{center}

{\Large\bf Gauge Kinetic Mixing and Leptophobic $Z'$ in $E_6$ and $SO(10)$}
\footnote{Work supported by the Department of 
Energy, Contract DE-AC03-76SF00515}
\medskip

\normalsize 
{\large Thomas G. Rizzo } \\
\vskip .3cm
Stanford Linear Accelerator Center \\
Stanford CA 94309, USA\\
\vskip .3cm

\end{center}

\begin{abstract} 
We examine the influence of gauge kinetic mixing on the couplings of a TeV 
scale $Z'$ in both $E_6$ and $SO(10)$ models. The strength of such mixing, 
which arises due to the existence of incomplete matter representations at low 
scale, can be described by a single parameter, $\delta$. The value of this 
parameter can significantly influence the ability of both hadron and lepton 
colliders to detect a $Z'$ using conventional search techniques. In addition, 
$\delta \neq 0$ also adds to the complexities involved in separating $E_6$ 
$Z'$ models from those arising from alternative scenarios. Employing a 
reasonable set of assumptions we have determined the allowed range for this 
parameter within a wide class of models via an RGE analysis. In particular, 
given the requirements of Standard Model gauge coupling unification, anomaly 
freedom and perturbativity up to the GUT scale, we demonstrate that the 
necessary condition for exact leptophobia in $\eta$ type $E_6$ models, 
$\delta=-1/3$, is impossible to achieve in this scenario. Furthermore we show 
that the allowed range for $\delta$ is rather restricted for arbitrary values 
of the mixing between the $U(1)_\chi$ and $U(1)_\psi$ type couplings. The 
$SO(10)$ $Z'$ model $\chi$ is discussed as a separate case since it requires 
special attention.
\end{abstract} 




\renewcommand{\thefootnote}{\arabic{footnote}} \end{titlepage}


\section{Introduction}

Though the Standard Model(SM) does an excellent job at describing precision 
electroweak data{\cite {moriond98}} there are many reasons to believe that 
new physics must exist at a scale not far above that which is currently being 
probed at colliders. The minimal supersymmetric extension of the SM, the 
MSSM, provides a setting for addressing a number of important questions left 
unanswered by the SM framework. Although convenient for many analyses due to 
its relative simplicity, no one truly expects that the MSSM will represent the 
actual version of SUSY realized by nature at low energies. Perhaps one of 
the simplest and well motivated extensions of the MSSM scenario is the 
enlargement of the SM gauge group, $SU(3)_C\times SU(2)_L\times U(1)_Y$, by 
additional $SU(2)$ or $U(1)$ factors. From the GUT or string point of view, 
the presence at low energies, $\sim  1$~TeV, of an additional neutral gauge 
boson, $Z'$, associated with a $U(1)'$ seems reasonably likely{\cite {newz}}. 
At a high energy scale, such a $Z'$ could arise naturally from, for example, 
the breaking of real or ersatz GUT such as $SO(10)$ or $E_6$ via patterns 
such as $SO(10)\to SU(5)\times U(1)_\chi$ or $E_6\to SO(10)\times U(1)_\psi$, 
with some linear combination of the $U(1)$'s surviving unbroken down to the 
TeV scale.

If such particles are indeed present they must either be reasonably massive, 
have small mixings with the SM $Z$, and/or have `unlucky' combinations of 
fermionic couplings in order to avoid direct searches at the 
Tevatron{\cite {cdfd0}} and potential conflict with precision electroweak 
data{\cite {hagi}}. One `unlucky' set of couplings 
that has gotten much attention in the literature is the condition know as 
leptophobia{\cite {leptop}}, \ie, where the $Z'$ does not couple to SM leptons. 
In such a situation the $Z'$ avoids traditional collider searches since it 
cannot be produced in Drell-Yan collisions and it does not perturb any of the 
leptonic coupling data collected at LEP through asymmetry measurements or 
the value of $A_{LR}$ obtained by SLD. To 
discover such a $Z'$ at the Tevatron or LHC would require the observation of 
a bump in the dijet mass spectrum, a difficult prospect due to large QCD 
backgrounds{\cite {bump}} and finite jet energy resolution. In the absence of 
mixing with the SM $Z$, the $Z'$ would also not be produced at future lepton 
colliders except via loops.

As is easily demonstrated, the condition of leptophobia does not exist in 
{\it conventional} $SO(10)$ or $E_6$ models{\cite {us}} where the fermionic 
couplings are essentially determined by group theory, the choice of embedding 
and, in the $E_6$ case, 
by the value of a mixing angle $\theta$. Interestingly, in the flipped 
(non-$SO(10)$ unified) 
$SU(5)\times U(1)_X${\cite {flip}} model, leptophobia is possible if one 
assumes that leptons do not carry $X$ quantum numbers (\ie, only the three 
{\bf 10}'s  
carries a non-zero $X$ charge) and one allows the $X$ charge assignments to 
be generation dependent in order to cancel anomalies. If we 
also demand that the $Z'$ couplings be at least approximately flavor diagonal 
in order to avoid problems associated with flavor changing neutral currents 
then there is no leptophobic $Z'$ case in this scheme as well. Of course it 
is always possible to directly construct leptophobic $Z'$ models with 
generation independent couplings following a purely phenomenological 
approach{\cite {hinch}} but it is not clear how such models are embedded in 
a larger framework.

In a recent series of papers, Babu, Kolda and March-Russell{\cite {babu}} 
discussed the possibility of constructing a leptophobic $Z'$ model within 
$E_6$-type models through the dynamical effects 
associated gauge boson kinetic mixing(KM){\cite {kinmix}} which occurs 
naturally at some level in almost all realistic GUT or string models. KM 
essentially arises due to the existence of incomplete GUT representations at 
the low energy scale. For example, such a situation is seen to occur 
even in the MSSM where the usual two Higgs doublet superfields are low energy 
survivors associated with part of a pair of 
{\bf 5}$+\overline{\mbox{\bf 5}}$'s at the high scale. While the 
Higgs doublet components are light the remaining dangerous, color triplet 
isosinglet pieces are forced phenomenologically to remain heavy by 
proton decay constraints. Even if KM is naturally absent 
at the high scale, the partitioning of any of the multiplets will drive KM to 
be non-zero at the TeV scale via the renormalization group equations(RGEs). If 
there are enough low energy survivors from split multiplets with the correct 
quantum numbers, Babu \etal ~showed that the effects of KM on the $Z'$ 
couplings can be sufficiently large to obtain leptophobic conditions.

The purpose of the present paper is to make a broad survey of models 
associated with new $U(1)$ factors arising from $E_6$ and $SO(10)$ and to 
ascertain quantitatively the impact of KM on the corresponding $Z'$ 
couplings. Clearly if leptophobia is indeed possible a $Z'$ may be missed by 
present and future collider searches. We will show, 
subject to a reasonable set of assumptions, that the values of the parameters 
necessary for {\it complete} leptophobia, \ie, identically zero vector and 
axial vector leptonic couplings, cannot be achieved in these models. We also 
show that although KM has dramatic consequences for the $Z'$ couplings in these 
scenarios it will still remain possible to discover the $Z'$ at both hadron and 
lepton colliders via their leptonic couplings. In addition we will show that 
it will still be possible to distinguish a $Z'$ originating from 
$E_6$ (including KM) from, \eg, a $Z'$ originating from the Left-Right 
Symmetric Model{\cite {lrm}} once the $Z'$ couplings are measured with 
reasonable accuracy at future hadron and lepton colliders. For the $Z'$ 
arising in $SO(10)$ we will show that ambiguities in identification remain 
unless the $Z'$ is directly produced at a lepton collider. 

This paper is organized as follows: in Section 2 we set up our notation and 
review the essentials of kinetic mixing and $Z'$ couplings in general $E_6$ 
models which incorporate KM. The possibilities associated with alternative 
fermion embedding schemes are discussed. We explicitly show how the search 
reaches for such a $Z'$ would be altered by an arbitrary amount of KM at both 
the Tevatron and LHC. We also show the KM impact on the couplings themselves 
and the possible confusion that can arise when trying to determine the model 
from which the $Z'$ originated if KM effects were allowed to be arbitrarily 
large. The basic formulae needed in our later analysis are also supplied here 
at the one-loop level. In Section 3 we discuss our model building assumptions 
and numerically analyze the resulting 68 $E_6$ models and 134 $SO(10)$ models 
to which these assumptions naturally lead. We demonstrate that exact 
leptophobia does not occur in any of these models even though the overall 
effects of KM can be numerically substantial. The resulting allowed range of 
couplings are determined in all cases. The influence of kinetic mixing on the 
$Z'$ search reaches of the Tevatron and LHC within these models is 
also examined in detail as are a number of issues relating to $Z'$ 
identification. A 
summary and discussion as well as our conclusions can be found in Section 4.

\section{Notation, Background and Review of Kinetic Mixing}

Consider the Lagrangian for the electroweak part of the SM with the addition 
of a new $U(1)$ field which is decomposed in the following manner:
\begin{equation}
{\cal L}={\cal L}_{kin}+{\cal L}_{int}+{\cal L}_{SB}+{\cal L}_{SUSY}\,,
\end{equation}
where the most general form of ${\cal L}_{kin}$ is given by 
\begin{equation}
{\cal L}_{kin}=-{1\over {4}}W^a_{\mu\nu}W^{a\mu\nu}-{1\over {4}}
\tilde B^{\mu\nu}\tilde B_{\mu\nu}-{1\over {4}}
\tilde Z'^{\mu\nu}\tilde Z'_{\mu\nu}-{\sin \chi \over {2}}\tilde B_{\mu\nu}
\tilde Z'^{\mu\nu}\,,
\end{equation}
with $W^a,\tilde B$ and $\tilde Z'$ representing the usual $SU(2)_L$, $U(1)_Y$ 
and $U(1)'$ fields, with the index `a' labelling the weak isospin. Note 
that the term proportional to $\sin \chi$ which 
directly couples the $\tilde B$ and $\tilde Z'$ fields is not forbidden by 
either $U(1)_Y$ or $U(1)'$ gauge invariance and corresponds to gauge 
kinetic mixing. In this basis the interaction terms for fermions can be 
written as 
\begin{equation}
{\cal L}_{int}=-\bar \psi \gamma^\mu[g_LT^aW^a_\mu+\tilde g_Y Y \tilde B_\mu 
+\tilde g_{Q'}Q'\tilde Z'_\mu]\psi\,. 
\end{equation}
The parts of the Lagrangian describing symmetry breaking and the interactions 
of the SUSY partners are contained in terms ${\cal L}_{SB}+{\cal L}_{SUSY}$ and 
will not directly concern us here. We can remove the off-diagonal coupling of 
the $\tilde B$ and $\tilde Z'$ in the kinetic energy by making the field 
transformations:
\begin{eqnarray}
\tilde B_\mu &=& B_\mu-\tan \chi Z'_\mu\,, \nonumber \\
\tilde Z'_\mu &=& {Z'_\mu \over {\cos \chi}}\,. 
\end{eqnarray}
This diagonalizes the kinetic terms in ${\cal L}_{kin}$ and, 
making the corresponding transformation in the couplings: 
\begin{eqnarray}
g_Y &=& \tilde g_Y\,, \nonumber \\
g_{Q'} &=& {\tilde g_{Q'} \over {\cos \chi}}\,, \nonumber \\
g_{YQ'} &=& -\tilde g_Y \tan \chi\,,
\end{eqnarray}
allows the interaction term in the Lagrangian to be written in a more 
familiar form. The couplings are assumed to be `GUT' normalized in this basis 
since we will assume that complete representations exist at the high scale. 
Using the SM notation and normalization conventions, \ie, 
$Y\to \sqrt {3\over {5}} Y_{SM}$ and $g_Y\to \sqrt {5\over {3}} g'$ such that 
$Q_{em}=T_{3L}+Y_{SM}$, we obtain the more traditional appearing result 
\begin{equation}
{\cal L}_{int}=-\bar \psi \gamma^\mu[g_LT^aW^a_\mu+g' Y_{SM} B_\mu 
+g_{Q'}(Q'+\sqrt {3\over {5}}\delta Y_{SM})Z'_\mu]\psi\,,
\end{equation}
where $\delta\equiv g_{YQ'}/g_{Q'}$ and we immediately recognize the usual SM 
weak isospin and hypercharge coupling terms. Note that $\delta \neq 0$ 
requires $g_{YQ'}\neq 0$. Of course, for our purposes we 
must remember that all of the couplings in this term run with energy and are 
thus to be evaluated at the EW or TeV scale to make contact with experiment. 
Furthermore, recalling that $g_{Q'}$ is GUT normalized, the $Z'$ piece of this 
interaction can also be re-written to conform to more 
conventional{\cite {us,steve}} notation, \ie, 
\begin{equation}
{\cal L}(Z')_{int}=-\lambda {g_L\over {c_w}}\sqrt {5x_w\over {3}}
\bar \psi \gamma^\mu(Q'+\sqrt {3\over {5}}\delta Y_{SM})\psi Z'_\mu\,,
\end{equation}
with as usual $x_w=\sin^2 \theta_w=e^2/g_L^2$, $c_w=\cos \theta_w$ and 
$\lambda=g_{Q'}/g_Y$. Note that in this notation 
$\delta \cdot \lambda=-\tan \chi$. 
Assuming for our purposes that the $Z'$ arises from the symmetry breaking chain 
$E_6\to SO(10)\times U(1)_\psi \to SU(5)\times U(1)_\chi \times U(1)_\psi \to
SM\times U(1)_\theta$, one obtains $Q'=Q_\psi \cos \theta-Q_\chi \sin \theta$, 
where $\theta$ is the familiar $E_6$ mixing angle and the 
$Q_{\psi,\chi, \eta}$, the last being the appropriate combination for model 
$\eta$ which corresponds to $\theta =\tan^{-1} \sqrt {3/5} \simeq 37.76^o$, are 
given in Table 1 assuming the conventional particle embeddings. (In the 
$SO(10)$ case to be discussed later we simply set $\theta=-\pi/2$ which 
corresponds to the $\chi$ model.)

\begin{table*}[htbp]
\leavevmode
\begin{center}
\label{numbers}
\begin{tabular}{lccccc}
\hline
\hline
Particle & $SU(3)_c$ & $2\sqrt{6} Q_\psi$ & $2\sqrt{10} Q_\chi$ & 
$2\sqrt{15} Q_\eta$ & Y \\
\hline
$Q=(u,d)^T$   & {\bf 3}      &   1   & -1   & 2 & 1/6    \\
$L=(\nu,e)^T$ & {\bf 1}      &   1   &  3   & -1 &-1/2   \\
$u^c$ &$\overline{\mbox{\bf 3}}$       &   1   &  -1  & 2& -2/3   \\
$d^c$ &$\overline{\mbox{\bf 3}}$       &   1   &  3   & -1 &1/3    \\
$e^c$         &{\bf 1}       &   1   &  -1  &  2 &1     \\
$\nu^c$       &{\bf 1}       &   1   &  -5  &  5 &0     \\
$H=(N,E)^T$   &{\bf 1}       &  -2   &  -2  &  -1 &-1/2  \\
$H^c=(N^c,E^c)^T$ &{\bf 1}   &  -2   &  2   &  -4 &1/2   \\
$h$           & {\bf 3}      &  -2   &  2   &  -4 &-1/3  \\
$h^c$ &$\overline{\mbox{\bf 3}}$       &  -2   &  -2  & -1 &   1/3  \\
$S^c$     &{\bf 1}           &   4   &  0   &  5 &0     \\
\hline
\hline
\end{tabular}
\caption{Quantum numbers of the particles contained in the {\bf 27} 
representation of $E_6$; standard particle embeddings are assumed and all 
fields are taken to be left-handed.}
\end{center}
\end{table*}

As we will see below, the a priori unknown parameters $\delta$ and $\lambda$ 
are directly calculable for any value of $\theta$ from an RGE analysis within 
the framework of a given model with fixed matter content assuming high scale 
coupling unification. Allowing both of these parameters to vary 
freely clearly leads to significant modifications of the potential $Z'$ 
couplings. However, as was noted by Babu \etal, if we do indeed treat them 
as free parameters one finds that for conventional particle embeddings with 
$\delta=-1/3$ and $\theta=\tan^{-1}\sqrt {3/5}$, 
\ie, the couplings of model $\eta$, both the vector and axial vector leptonic 
couplings of the $Z'$ vanish for all values of $\lambda$ and leptophobia is 
obtained. A quick analysis shows that this choice of parameters 
is {\it unique}. In alternative embeddings leptophobia is also 
possible but its location in the model parameter space is modified. In the 
case of the flipped $SU(5)$-type model {\cite {flip}}, the roles played by 
the pairs $(u^c,e^c)$ and $(d^c,\nu^c)$ are interchanged, so that the 
lepton's right-handed couplings are modified. Similarly, in the alternative 
embedding scheme of Ma{\cite {ma}}, the fields $(L,d^c,\nu^c)$ are 
interchanged with $(H,h^c,S^c)$, which also leads to leptonic coupling 
changes, this time for the left-handed couplings. Of course we 
can also imagine both interchanges being made simultaneously leading to a 
fourth set of possible leptonic couplings. In each of these cases a unique 
point in the $\theta-\delta$ parameter space leads to leptophobia; these are 
summarized in Table 2. Note that both the standard and the alternative 
embedding due to Ma lead to the same required values of the parameters in 
order to achieve leptophobia. This is not too surprising as model $\eta$ 
couplings are invariant under the particle interchange associated with Ma's 
model. In all cases we see that 
the required magnitude of $\delta$ to achieve leptophobia is reasonably large. 
We also note from this Table that the $SO(10)$-inspired $\chi$ model can never 
be even approximately leptophobic independently of how the particles are 
embedded. 

\begin{table*}[htbp]
\leavevmode
\begin{center}
\label{leptoloc}
\begin{tabular}{lcc}
\hline
\hline
Embedding & $\tan \theta$ & $\delta$ \\
\hline
Standard &$\sqrt {3/5}$   & $-1/3$   \\
Flipped  &$\sqrt {15}$    &  $-\sqrt {10}/3$ \\
Ma       &$\sqrt {3/5}$   & $-1/3$   \\
Both     &$\sqrt {5/27}$  & $-\sqrt {5/12}$  \\
\hline
\hline
\end{tabular}
\caption{ Values of the parameters $\theta$ and $\delta$ for which exact 
leptophobia is obtained for the various embedding schemes discussed in the 
text.}
\end{center}
\end{table*}

\begin{figure}[htbp]
\centerline{
\psfig{figure=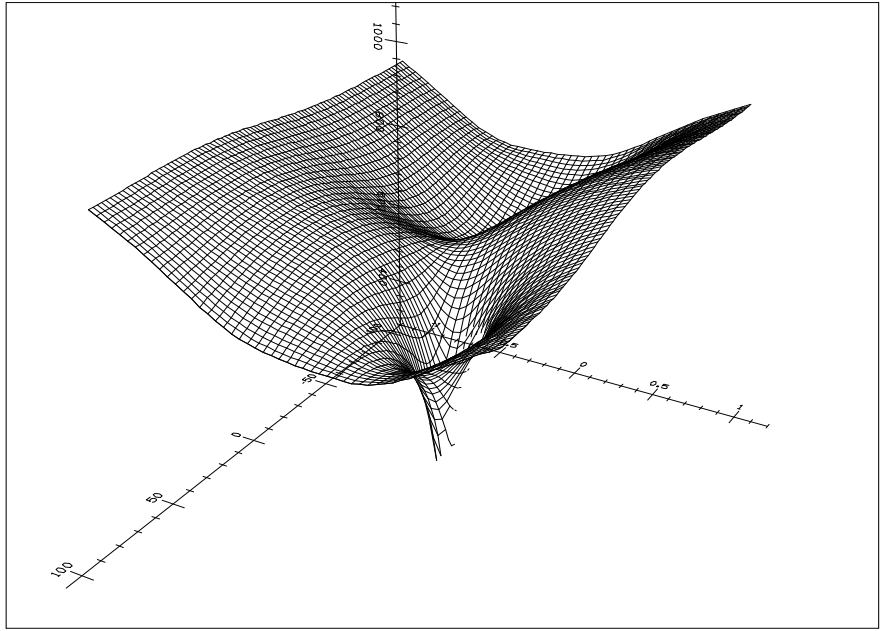,height=11.9cm,width=17cm,angle=-90}}
\vspace*{-28mm}
\centerline{
\psfig{figure=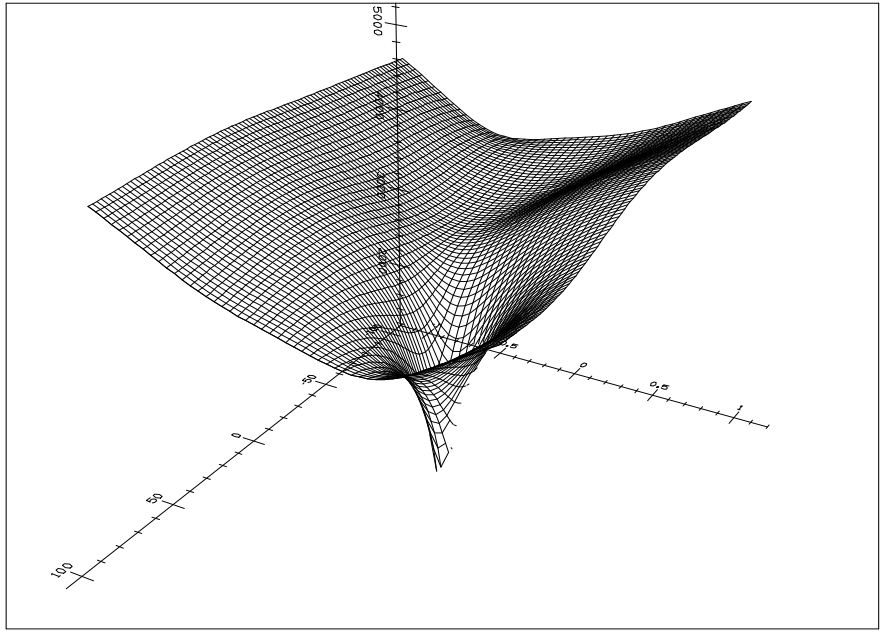,height=11.9cm,width=17cm,angle=-90}}
\vspace*{-1.05cm}
\caption{Search reaches for the $E_6$ $Z'$ at (top) the Tevatron($2~fb^{-1}$)  
and(bottom) LHC($100~fb^{-1}$) in GeV as functions of $\theta$(left axis) and 
$\delta$(right axis)  
assuming no exotic decay modes. The leptophobic hole is evident in both cases. 
The sign of $\delta$ has been reversed in these plots for ease of viewing and 
$\lambda=1$ has been assumed.}
\label{fig1}
\end{figure}

To get an idea of the potential impact of leptophobia, and $\delta \neq 0$ 
in general, we show in Fig.~\ref{fig1} the search reaches for an $E_6$ $Z'$, 
assuming the canonical particle embedding and assuming that the $Z'$ 
decays only to SM particles, at both the Tevatron Run II and the LHC; we take 
the value $\lambda=1$ and use the CTEQ4M 
parton densities{\cite {cteq}}. (For other values of 
$\lambda$ near unity the mass reaches scale approximately as 
$\Delta M \simeq 180 \log (\lambda)$ GeV and 
$\Delta M \simeq 660 \log (\lambda)$ GeV at TeV II and LHC, respectively.) 
In both 
cases the reach is roughly $\theta$ and $\delta$ independent ($\simeq 850$ GeV 
and $\simeq 4200$ GeV for TeV II and the LHC, respectively) except near the 
leptophobic region where it falls off quite dramatically forming a hole in 
the mass reach. It 
is clear that the conventional $Z'$ searches will fail in this region and that 
the dijet method would need to be employed to find the $Z'$. 

\vspace*{-0.5cm}
\nn
\vspace*{0.4mm}

\begin{figure}[htbp]
\centerline{
\psfig{figure=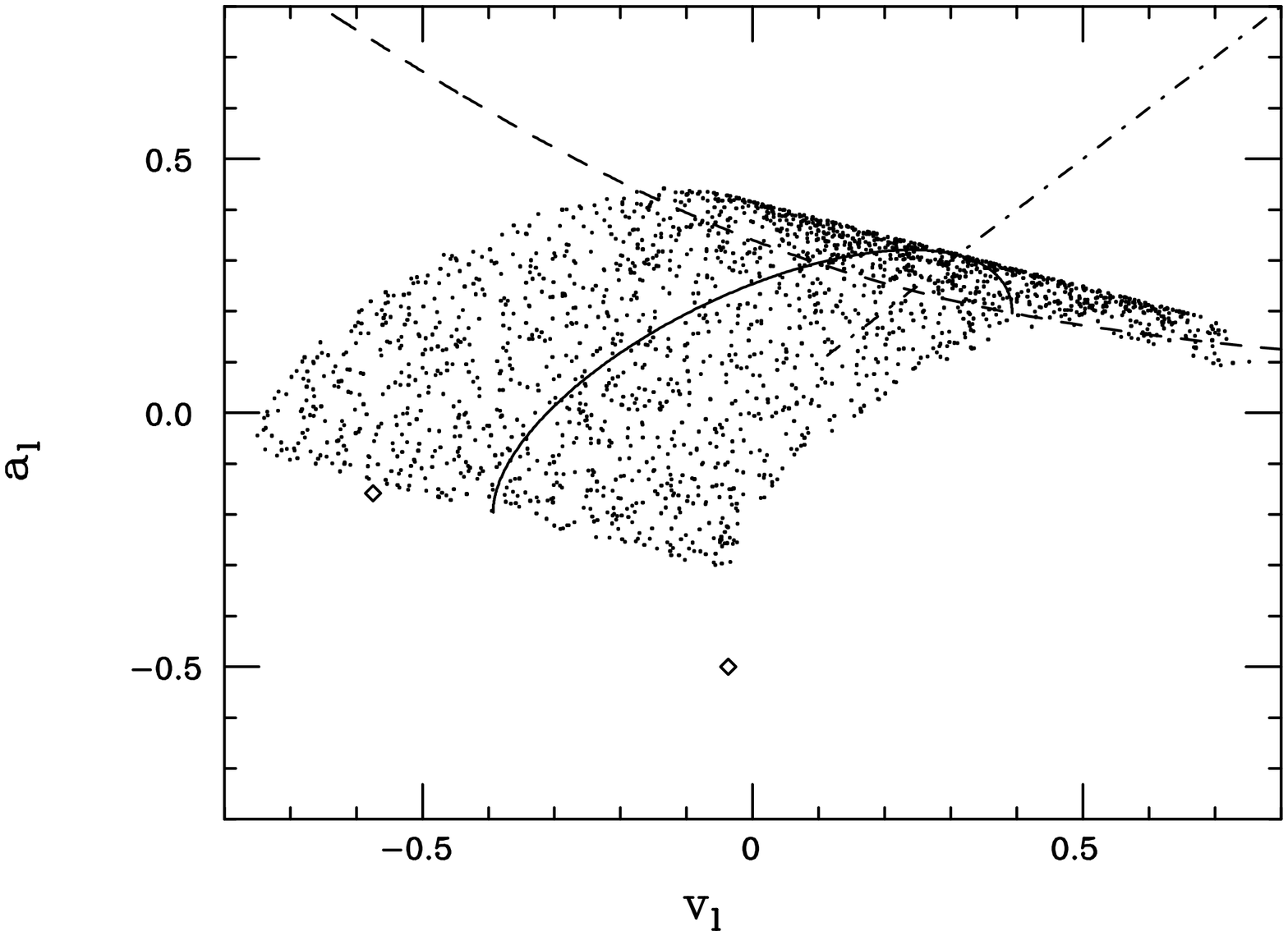,height=10.5cm,width=13cm,angle=-90}}
\vspace*{-10mm}
\centerline{
\psfig{figure=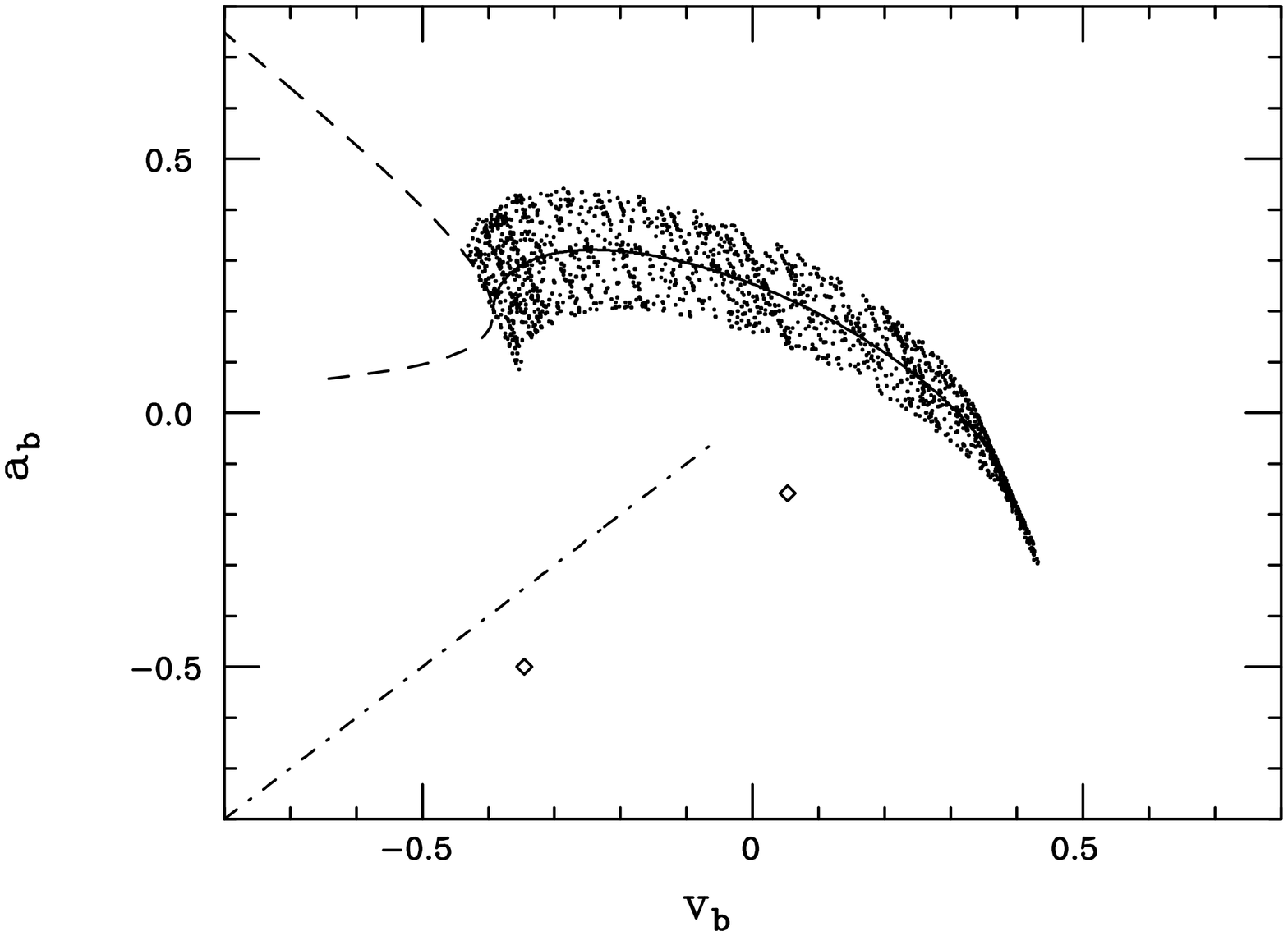,height=10.5cm,width=13cm,angle=-90}}
\vspace*{-0.9cm}
\caption{Vector and axial vector couplings for leptons(top) and 
$b$-quarks(bottom) in various $Z'$ models: the $E_6$ model with no KM(solid), 
the Left-Right Model(dashed), and the Un-unified Model(dash-dot), 
as well as the case of a heavy SM $Z'$ and the Alternative Model 
of Ma(labeled by the two diamonds.) The points are the predicted 
values in $E_6$ with KM assuming $-1/2 \leq \delta \leq 1/2$ and $\lambda=1$. 
For $\lambda \neq 1$ the predicted coupling region scales appropriately.}
\label{fig2}
\end{figure}

Arbitrarily large values of $\delta$ can also lead to possible confusion 
when $Z'$ 
couplings are extracted at, \eg, future lepton colliders. It is well known 
that when KM is absent sufficient data on $Z'$ couplings can be extracted at 
such machines, even below the $Z'$ production threshold, so that the $Z'$'s 
model of origin can be identified{\cite {snow,steve}}. When $\delta \neq 0$, 
the allowed ranges of the various vector and axial vector fermionic couplings 
in $E_6$ models is greatly extended in comparison to the more conventional 
case creating overlaps with the corresponding 
coupling values anticipated in other models. 
This result is shown explicitly in Fig.~\ref{fig2} for both leptonic and $b$ 
quark couplings, these being the ones most easily measured. Here the 
$E_6$ case with and without KM is compared to the 
predictions of the Left-Right Model{\cite {lrm}}, Ma's Alternative 
model{\cite {ma}}, the Un-unified 
Model{\cite {uum}} as well as to the reference case of a $Z'$ with SM 
couplings. In the KM case it has been assumed for simplicity that 
$\lambda=1$ and $\delta$ is confined to the range 
$-1/2 \leq \delta \leq 1/2$. One sees immediately that the 
presence of KM leads to potential mis-identification of the $Z'$ even when 
precise measurements of the couplings are available. Though not shown, 
similar effects would be observed in $u$-quark type couplings. Clearly, if 
the range of $\delta$ is increased and/or 
$\lambda$ were allowed to vary from unity by as small a value as say $25\%$, 
the size of the $E_6$ coupling region would dramatically increase and the $Z'$ 
mis-identification potential would rise dramatically. Note that the $SO(10)$ 
inspired $\chi$ model in the absence of KM corresponds to the point of contact 
of the solid and dashed curves, \ie, the non-KM $E_6$ and LRM cases, in both 
the coupling planes. In this case the $Z'$ in the non-KM $\chi$ model has the 
same couplings as does the $Z'$ in the LRM with 
$\kappa^2=(g_R/g_L)^2={5\over {3}} {x_w\over {(1-x_w)}}$, with $g_{L,R}$ 
being the gauge coupling associated with the $SU(2)_{L,R}$ group factor. 
From this analysis it is 
clear that apart from the specific problems of leptophobia it is very 
important to determine what the allowed ranges of both $\delta$ and 
$\lambda$ are in realistic $E_6$ and $SO(10)$ models.

\vspace*{-0.5cm}
\nn
\vspace*{0.4mm}

In order to constrain the low scale values of both $\lambda$ and $\delta$ for 
a given model we must first perform an RGE analysis. 
The coupled RGEs for $g_Y$, $g_{Q'}$ and $g_{YQ'}$ at one-loop are given 
in our notation by{\cite {babu}} 
\begin{eqnarray}
{dg^2_Y\over {dt}} &=& {(g^2_Y)^2\over {8\pi^2}}B_{YY}\,, \nonumber \\
{dg^2_{Q'}\over {dt}} &=& {g^2_{Q'}\over {8\pi^2}}[g^2_{Q'} B_{Q'Q'}+g^2_{YQ'}
B_{YY}+2g_{Q'}g_{YQ'}B_{YQ'}]\,, \\
{dg^2_{YQ'}\over {dt}} &=& {1\over {8\pi^2}}[g^2_{Q'}g^2_{YQ'}B_{Q'Q'}+g^4_{YQ'}
B_{YY}+2g^2_Yg^2_{YQ'}B_{YY}+2g^2_Yg_{Q'}g_{YQ'}B_{YQ'}+2g_{Q'}g^3_{YQ'}
B_{YQ'}]\,, \nonumber 
\end{eqnarray}
where $B_{ij}=Tr(Q_iQ_j)$, with the trace extending over the full low energy 
matter spectrum. In particular, $B_{YY}=\beta_Y={3\over {5}}Tr(Y^2)$ 
is the conventional GUT normalized beta function for the $U(1)_Y$ coupling. 
At the high (GUT or string) scale where complete multiplets are present one 
finds that $B_{YQ'}=0$, identically, so that $g_{YQ'}$, and hence, 
$\delta = g_{YQ'}/g_{Q'}$=0. Below the high scale we imagine that at least 
some {\it incomplete} matter multiplets survive to low energies rendering 
$\delta \neq 0$ via RGE evolution. The quantum numbers of these survivors will 
tell us the specific value of $\delta$. It is important to stress that 
$B_{YQ'}$ receives no contributions from complete multiplets or from SM 
singlets.

Since gauge invariance tells us that there is no mixing between the 
$SU(3)_C$, $SU(2)_L$ and either of the $U(1)$ gauge fields, the one-loop RGEs 
for both the $g_{L,s}$ couplings take their conventional forms and can be 
trivially analytically integrated. Writing $L=\log (M_U/M_Z)$, these two 
equations can be combined as usual from which we obtain 
\begin{eqnarray}
L &=& {{2\pi (\alpha^{-1}_s-x_w\alpha^{-1})}\over {\beta_s-\beta_L}}
\,, \nonumber \\
\alpha^{-1}_U  &=&  {{\beta_s x_w \alpha^{-1}-\beta_L\alpha^{-1}_s}
\over {\beta_s-\beta_L}} \,, 
\end{eqnarray}
with $\alpha_U$ being the common unification coupling. Further, integration 
of the hypercharge RGE yields the usual result 
\begin{equation}
g^{-2}_Y(t)={\alpha^{-1}_U\over {4\pi}}[1+{\alpha_U \over {2\pi}} B_{YY}
(t_U-t)]\,,
\end{equation}
where $t_U \sim \log M_U$ and the unification boundary condition has been 
imposed. 

Since $\delta=g_{YQ'}/g_{Q'}$, and the solution for $g^2_{Y}(t)$ is known, we 
can combine the last two of the RGEs above to obtain
\begin{eqnarray}
{d\delta\over {dt}}&=&{1\over {g_{Q'}}}{dg_{YQ'}\over {dt}}-
{g_{YQ'}\over {g^2_{Q'}}}{dg_{Q'}\over {dt}}\,, \nonumber \\
     &=& {g^2_Y\over {8\pi^2}}[B_{YQ'}+\delta B_{YY}]\,.
\end{eqnarray}
This can now be directly integrated with the result
\begin{equation}
\delta(t)=-{B_{YQ'}\over {B_{YY}}}\big[1-[1+{\alpha_U B_{YY}(t_U-t)
\over {2\pi}}]^{-1}\big]\,,
\end{equation}
where we have imposed the boundary condition that 
$\delta(t)$ vanishes at the GUT scale $M_U$ since we assume that complete 
multiplets exist there. The weak scale parameter $\delta$ relevant for 
the $Z'$ couplings is obtained when we set $t\sim \log M_Z$ so that 
$t_u-t=\log (M_U/M_Z)=L$ in the expression above. Note that $\delta$ grows as 
the value of $B_{YQ'}$ increases. From this expression it is 
obvious that we need to have as many split multiplets as possible at low 
energies in order to enhance the value of $\delta$. Knowing $\delta(t)$ then 
allows us to re-write the RGE for $g^2_{Q'}$ as 
\begin{equation}
{dg^2_{Q'}\over {dt}} = {(g^2_{Q'})^2\over {8\pi^2}}[B_{Q'Q'}+2\delta(t)
B_{YQ'}+\delta^2(t)B_{YY}]\,,
\end{equation}
which also can be integrated analytically. Defining the combination 
$z=\alpha_UB_{YY}L/2\pi$ we find 
\begin{equation}
g^{-2}_{Q'}(M_Z)={\alpha^{-1}_U\over {4\pi}}+{B_{Q'Q'}L\over {8\pi^2}}
\big[1-{B^2_{YQ'}\over {B_{YY}B_{Q'Q'}}}{z\over {1+z}}\big]\,,
\end{equation}
from which the coupling strength parameter $\lambda$ can be immediately 
calculated. We are now set to examine the values of $\delta$ and $\lambda$ 
that can arise in a given model.

\section{Models and Results}

In order to proceed we must consider how the low energy particle content of 
our models is to be chosen. These will follow from the following set of 
basic model building assumptions:

\begin{itemize}

\item  The SM gauge couplings, together with that of the new $U(1)'$, are 
assumed to perturbatively unify at a high scale as in the MSSM. This has two 
immediate consequences: ($i$) we can add only sets of particles that would 
form complete multiplets under $SU(5)$, at least as far as their SM quantum 
numbers are concerned. ($ii$) The number and types of new fields is restricted 
since perturbative unification is lost if too many 
multiplets{\cite {pu}} are added. 

\item  All anomalies including those associated with the new $U(1)'$ must 
cancel amongst the low energy matter fields in the model. 

\item  Additional matter multiplets beyond those contained in the MSSM must 
be vector-like with respect to (at least) the SM. This not only helps with 
the anomaly problem but allows these new light fields not to make too large 
of a contribution to the oblique parameters{\cite {obl}} forcing a conflict 
with precision electroweak data. When combined with the above requirements 
this tells us that at low energies we may add at most four 
${\bf 5}+\overline{\mbox{\bf 5}}$'s or one 
${\bf 5}+\overline{\mbox{\bf 5}}$ plus one 
${\bf 10}+\overline{\mbox{\bf 10}}$, in addition to $SU(5)$ singlets, to the 
MSSM spectrum. All higher dimensional representations are excluded. 
Note that the addition of SM or $SU(5)$ singlets will leave 
$\delta$ invariant since neither $B_{YY}$ or $B_{YQ'}$ will be changed. 
However, $B_{Q'Q'}$ is altered in this case leading to a shift 
in the value of $\lambda$. 

\item  The new matter fields are assumed to be low energy survivors from 
either ${\bf 27}+\overline{\mbox{\bf 27}}$'s or from {\bf 78}'s of $E_6$ 
since these are automatically anomaly free even under the full $E_6$ gauge 
group and may arise from strings. 

\end{itemize}

Given this set of conditions we can consider a number of specific cases 
beginning with $E_6$ itself.

\subsection{$E_6$}

Here we know that the low energy theory contains three {\bf 27}'s as well as 
a pair of `Higgs' doublets, which we label as $H_1,H_1^c$ to avoid confusion 
with the members of the {\bf 27}, as in the MSSM. Complete {\bf 27}'s are 
necessary so that the $U(1)_\theta$ anomalies cancel for arbitrary values of 
$\theta$. (The case $\theta=-90^o$ corresponding to the $Z'$ from $SO(10)$ 
will be discussed separately in the next subsection.) These `Higgs' are then 
the minimal split multiplet content at low energies. 
(`Higgs' is here in quotes as we really mean a pair of superfields with 
Higgs-like quantum numbers which may or may not obtain vacuum expectation 
values. In principle some combination of the fields $H_1/H_1^c$ and those 
in the {\bf 27} will play the role of the Higgs doublets in the MSSM.)  
As was pointed out early on, the theory without these extra `Higgs' 
and only the $H/H^c$ components of the {\bf 27}'s responsible for spontaneous 
symmetry breaking, will not unify{\cite {hrr}}. These `Higgs' must arise from 
either a 
${\bf 27}+\overline{\mbox{\bf 27}}$ or {\bf 78} to avoid anomalies. Since the 
three {\bf 27}'s already contain three pairs of 
${\bf 5}+\overline{\mbox{\bf 5}}$ in addition to singlets in comparison to the 
MSSM, we are free {\it at most} 
to only add a single (ersatz) ${\bf 5}+\overline{\mbox{\bf 5}}$ to the low 
energy spectrum. Since the 
$H_1,H_1^c$ fields also originate from a ${\bf 5}+\overline{\mbox{\bf 5}}$ it 
is necessary to examine the $U(1)_{\psi,\chi}$ quantum numbers of these 
additional fields since this is all that distinguishes amongst them.  The 
${\bf 27}+\overline{\mbox{\bf 27}}$ contains 
three different choices: (1)~${\bf 5}(-2,2)+\overline{\mbox{\bf 5}}(2,-2)$, 
(2)~${\bf 5}(2,2)+\overline{\mbox{\bf 5}}(-2,-2)$ and 
(3)~${\bf 5}(-1,-3)+\overline{\mbox{\bf 5}}(1,3)$, where the numbers 
in the parentheses are the $Q_{\psi,\chi}$ quantum numbers as normalized in 
Table 1. The {\bf 78} on the otherhand contains only 
one candidate (4)~${\bf 5}(3,-3)+\overline{\mbox{\bf 5}}(-3,3)$; this last 
case corresponds to the field content of the `minimal' model presented by 
Babu \etal ~when $\eta$-type couplings are assumed{\cite {babu}}. For each of 
these cases the corresponding contributions to $B_{YQ'}$ and $B_{Q'Q'}$ can 
immediately written down. (The contribution of the three {\bf 27}'s to 
$B_{Q'Q'}$ is 9, independently of $\theta$.) For example, defining 
$a=\cos \theta /(2\sqrt 6)$ and $b=\sin \theta /(2\sqrt 10)$ one easily 
obtains the results for the $H_1/H_1^c$ fields for each of the cases (1)-(4) 
is given by
\begin{eqnarray}
B_{YQ'}(1) &=& -4 \sqrt {3\over {5}}(a+b)\,, \nonumber \\
B_{YQ'}(2) &=& 4 \sqrt {3\over {5}}(a-b)\,, \nonumber \\
B_{YQ'}(3) &=& 2 \sqrt {3\over {5}}(-a+3b)\,, \nonumber \\
B_{YQ'}(4) &=& 6 \sqrt {3\over {5}}(a+b)\,, 
\end{eqnarray}
and, correspondingly, 
\begin{eqnarray}
\Delta B_{Q'Q'}(1) &=& 16(a^2+b^2+2ab)\,, \nonumber \\
\Delta B_{Q'Q'}(2) &=& 16(a^2+b^2-2ab)\,, \nonumber \\
\Delta B_{Q'Q'}(3) &=& 4(a^2+9b^2-6ab)\,, \nonumber \\
\Delta B_{Q'Q'}(4) &=& 36(a^2+b^2+2ab)\,.
\end{eqnarray}
The contribution of the color triplet pieces of the same 
${\bf 5}+\overline{\mbox{\bf 5}}$'s is identical for $B_{Q'Q'}$ and of 
opposites sign for $B_{YQ'}$.

\begin{figure}[htbp]
\centerline{
\psfig{figure=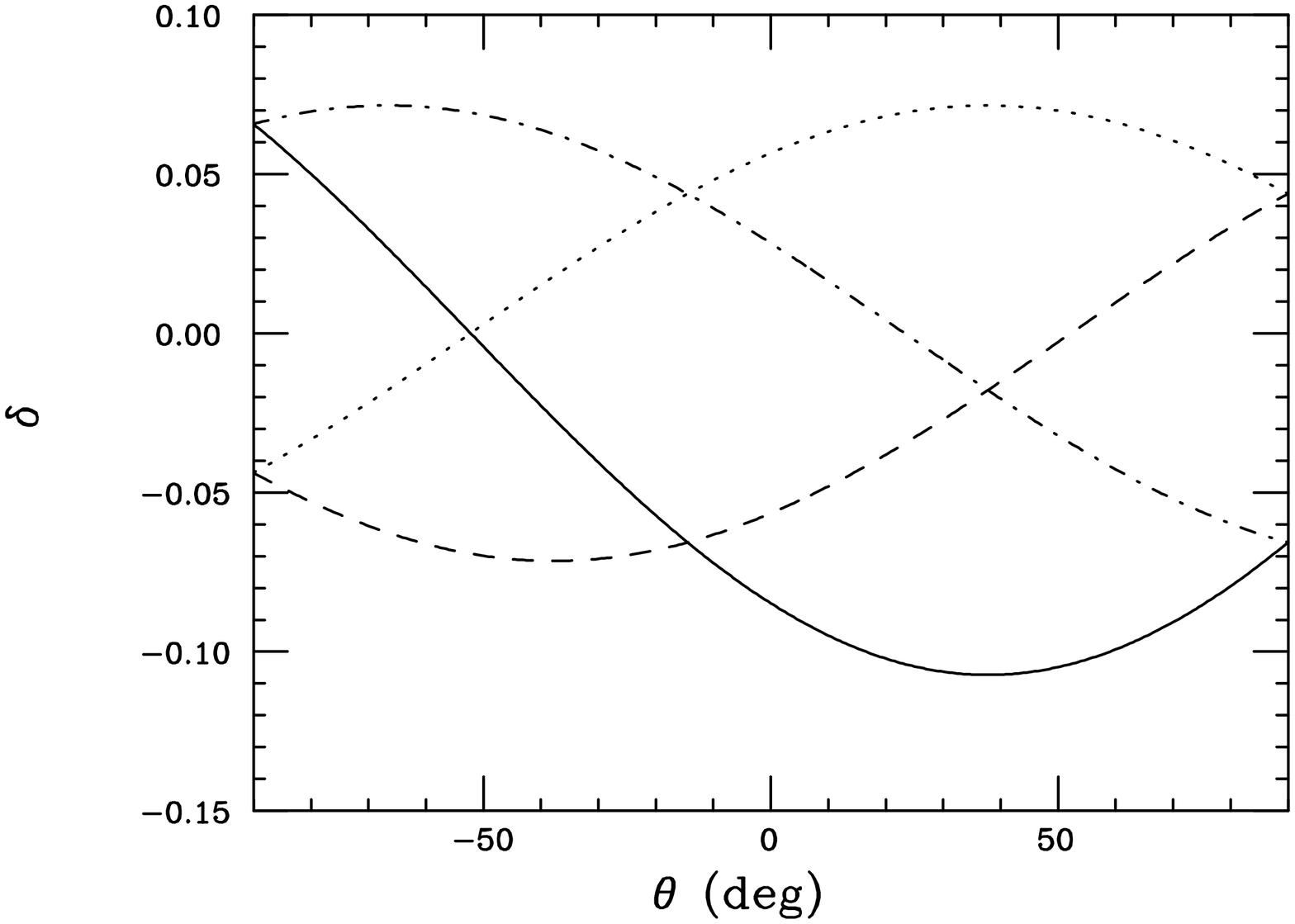,height=10.5cm,width=13cm,angle=-90}}
\vspace*{-5mm}
\centerline{
\psfig{figure=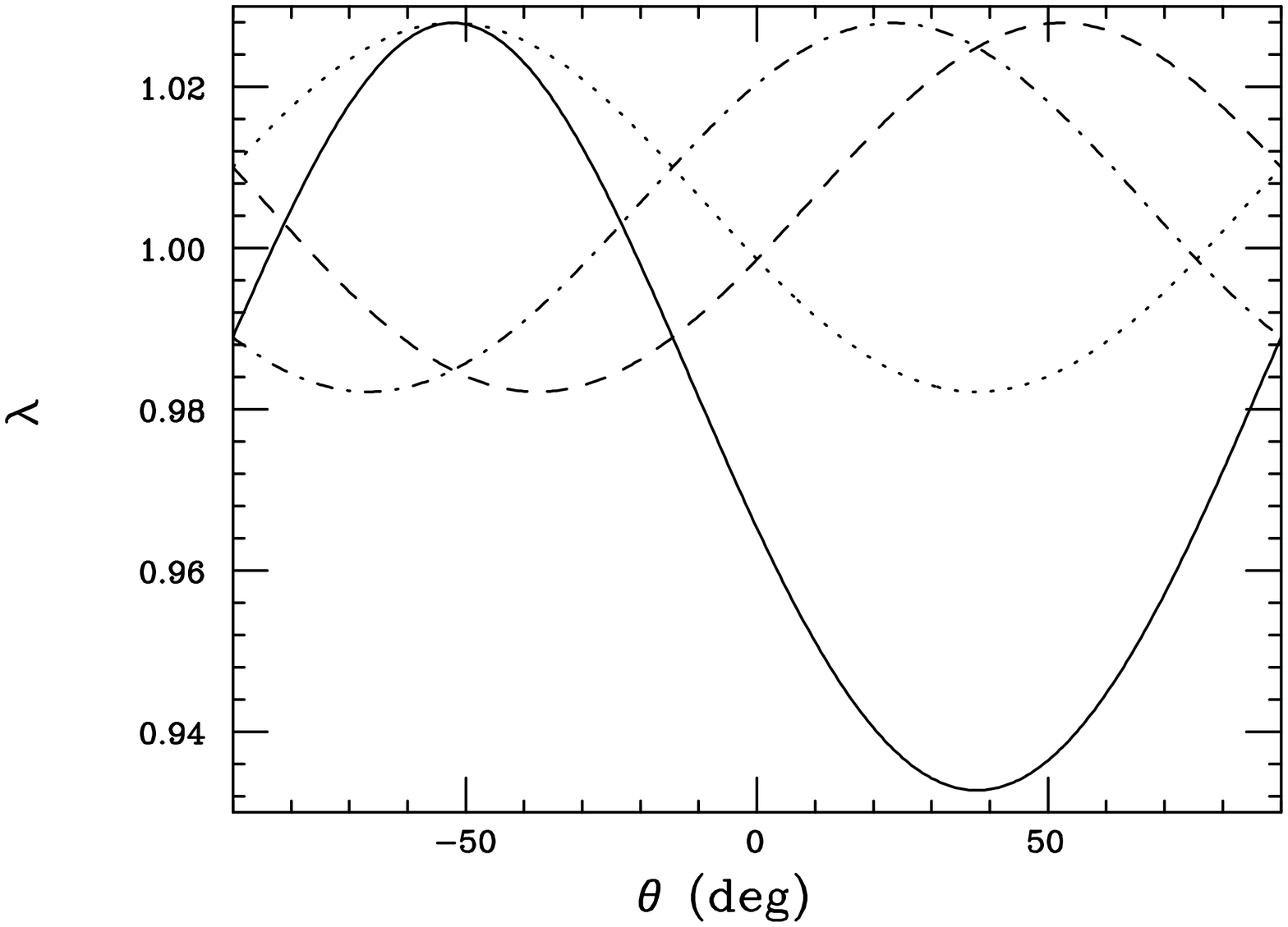,height=10.5cm,width=13cm,angle=-90}}
\vspace*{-0.9cm}
\caption{Predicted values of the parameters $\delta$(top) and 
$\lambda$(bottom) for four $E_6$ case ($i$) possibilities 
discussed in the text. The dotted, dashed, 
dash-dotted and solid curves correspond to the embedding 
choices (1)-(4), respectively.}
\label{fig3}
\end{figure}

As discussed above there are thus only two possible subcases to consider. 
Either ($i$) $H_1/H_1^c$ is the only pair of light superfields beyond the 
three {\bf 27}'s or ($ii$) the field content of an additional 
${\bf 5}+\overline{\mbox{\bf 5}}$ is also present. In case ($i$) we know 
immediately that $\beta_s=0$, $\beta_L=4$ and $B_{YY}={48\over {5}}$. The 
values of both $B_{Q'Q'}$ and $B_{YQ'}$ can also be directly calculated as 
above but depend visibly 
upon the choice, (1)-(4), into which we embed the 
$H_1/H_1^c$ fields as well as the value of $\theta$. With only 4 choices for 
the $H_1/H_1^c$ quantum numbers, the 
calculation is straightforward and we arrive at the results shown in 
Figure~\ref{fig3}. (For numerical purposes we have taken 
$\alpha_s(M_Z)=0.119${\cite {qcd}}, 
$\alpha_{em}^{-1}(M_Z)=127.935${\cite {erler}} and 
$\sin ^2 \theta_w=0.23149${\cite {moriond98}}; our results 
depend only weakly on these particular choices.) From the Figure several 
observations are immediate. First, both $\delta$ and $\lambda$ are 
constrained to rather narrow ranges and leptophobia is not obtainable. 
Second, the specific predicted values of $\delta$ and $\lambda$ depend 
quite sensitively on the embedding choices (1)-(4). Lastly, both $\delta$ and 
$\lambda$ are also strongly $\theta$ dependent but the choice of $\eta$ 
couplings, \ie, $\theta \simeq 37.76^o$,  extremizes their values. Since 
$\delta \to 0$ and $\lambda \to 1$ 
as we raise the survivor mass scale above $M_Z$, the curves actually represent 
the extreme boundaries of the parameter range obtainable for these quantities 
for case ($i$). Note that for $\eta$ couplings and $H_1/H_1^c$ embedding (4) 
we recover the value $\delta \simeq -0.11$ obtained{\cite {babu}} by 
Babu \etal ~in their so-called `minimal' model.

\vspace*{-0.5cm}
\nn
\vspace*{0.4mm}

In case ($ii$) the situation is somewhat more complex since the low energy 
spectrum now contains two `Higgs' doublets, $H_{1,2}/H_{1,2}^c$ as well as a 
pair of isosinglet, color triplet superfields, $D_1,D_1^c$. This uniquely 
fixes the values $\beta_s=1$, $\beta_L=5$ and $B_{YY}={53\over {5}}$ but 
allows for $4^3=64$ possible(but not necessarily independent), 
$\theta$-dependent values for $B_{YQ'}$ and $B_{Q'Q'}$. We can label our cases 
by the triplet (i,j,k) where the first(second,third) index labels the 
embedding choice, \ie, (1)-(4), for the field 
$H_1/H_1^c$($H_2/H_2^c$,~$D_1/D_1^c$). For example, we may choose $H_1/H_1^c$ 
to be from (1), $H_2/H_2^c$ from (3) and $D_1/D_1^c$ from (4) and we would 
label this subcase as (1,3,4). 
All of the contributions can be directly obtained from the last two equations 
by choosing appropriate combinations.

\vspace*{-0.5cm}
\nn
\begin{figure}[htbp]
\centerline{
\psfig{figure=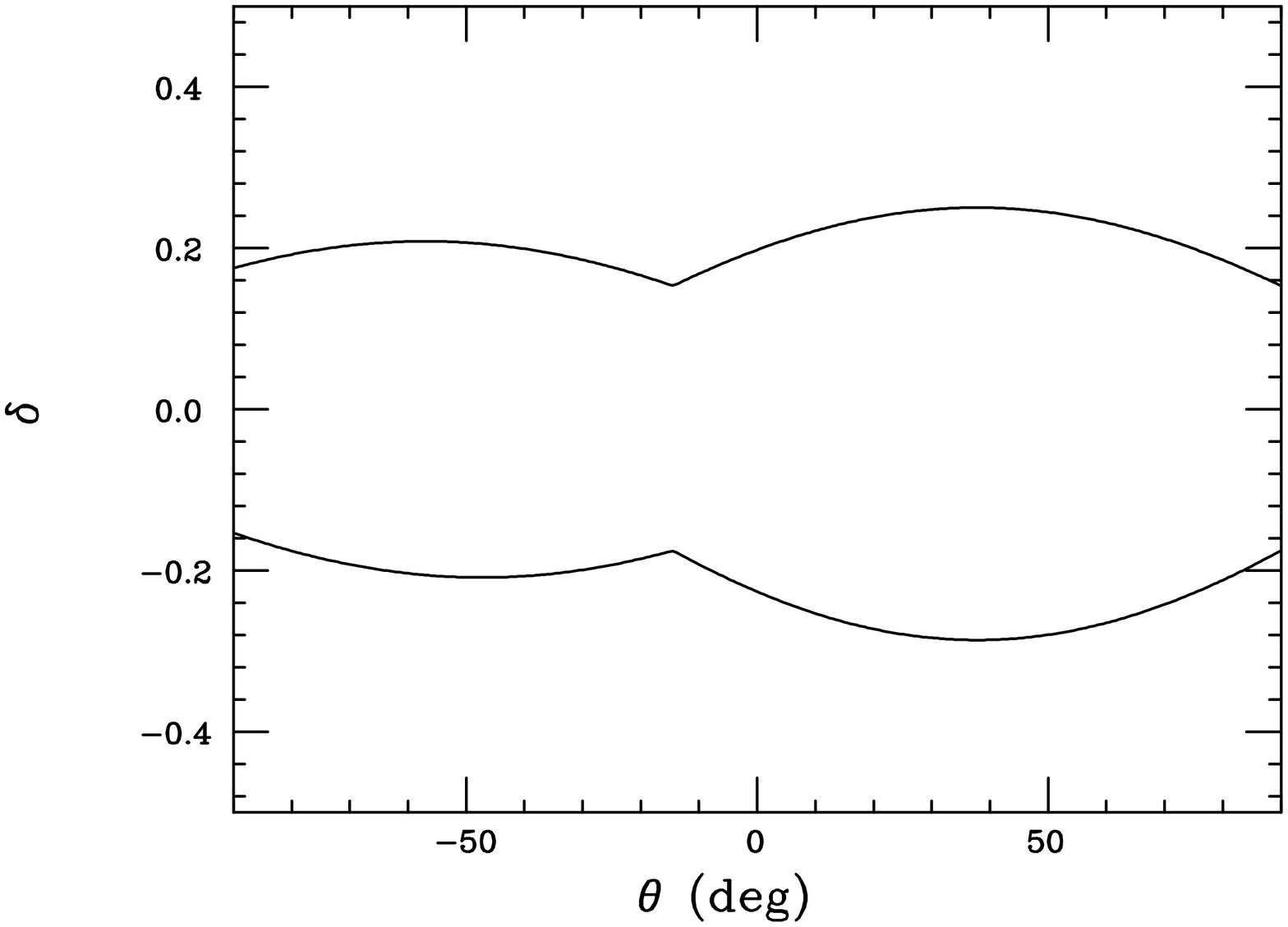,height=10.5cm,width=13cm,angle=-90}}
\vspace*{-5mm}
\centerline{
\psfig{figure=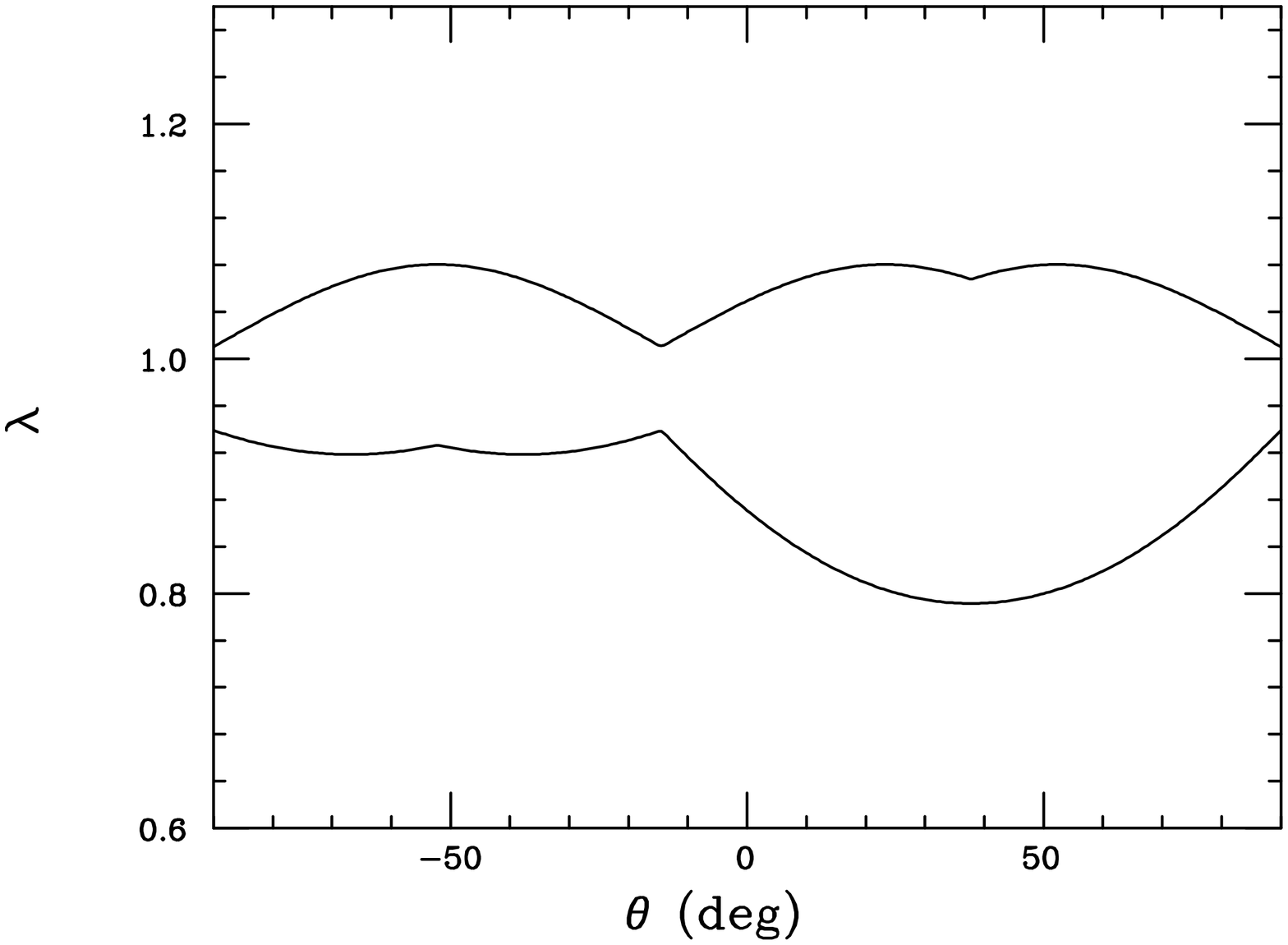,height=10.5cm,width=13cm,angle=-90}}
\vspace*{-0.9cm}
\caption{Boundaries of the allowed ranges for $\delta$(top) and 
$\lambda$(bottom) for the 64 case, type ($ii$) $E_6$ models discussed in the 
text as functions of the mixing angle $\theta$.}
\label{fig4}
\end{figure}
\vspace*{0.4mm}

We have calculated both $\delta$ and $\lambda$ for each of the 64 cases; 
Figure~\ref{fig4} shows the `envelope' of the range of values of $\delta$ 
and $\lambda$ as functions of $\theta$. In all cases the actual values must 
lie within the `envelope'. Several observations are possible from these 
results. First, independently of the value of $\theta$, we obtain the bounds 
$-0.286 \leq \delta \leq 0.250$ and $0.791 \leq \lambda \leq 1.080$ so that 
{\it exact} leptophobia is not achieved anywhere in the parameter space. 
(Just how close we are to leptophobia will be discussed below.)  The 
minimum[maximum] value of $\delta \simeq -0.286[+0.250]$ is achieved for 
$\eta$-type couplings with the embedding (4,4,1)[(1,1,4)] which corresponds 
to the so-called `maximal' model of Babu \etal {\cite {babu}}. The extrema for 
$\lambda$, \ie, $\lambda=0.791[1.080]$ are obtained for embeddings 
(4,4,4)[(1,1,1)] for $\eta$-type couplings and $\theta=-52.24^o$ (model I),  
respectively. Interestingly, the range of $\delta$ is sufficiently narrow so 
that none of the models listed in Table 2 can achieve leptophobic conditions. 
Next, we note that a 
further contribution to apparent leptophobia can occur in the $\eta$ coupling 
region since it is there that one obtains the smallest values of $\lambda$, 
rescaling the couplings to smaller values. 
(Recall, the Drell-Yan rate for the $Z'$ scales as $\lambda^2$.) It would be 
nice to perform a two-loop RGE calculation to verify these leading order 
results once these equations become available.

\vspace*{-0.5cm}
\nn
\begin{figure}[htbp]
\centerline{
\psfig{figure=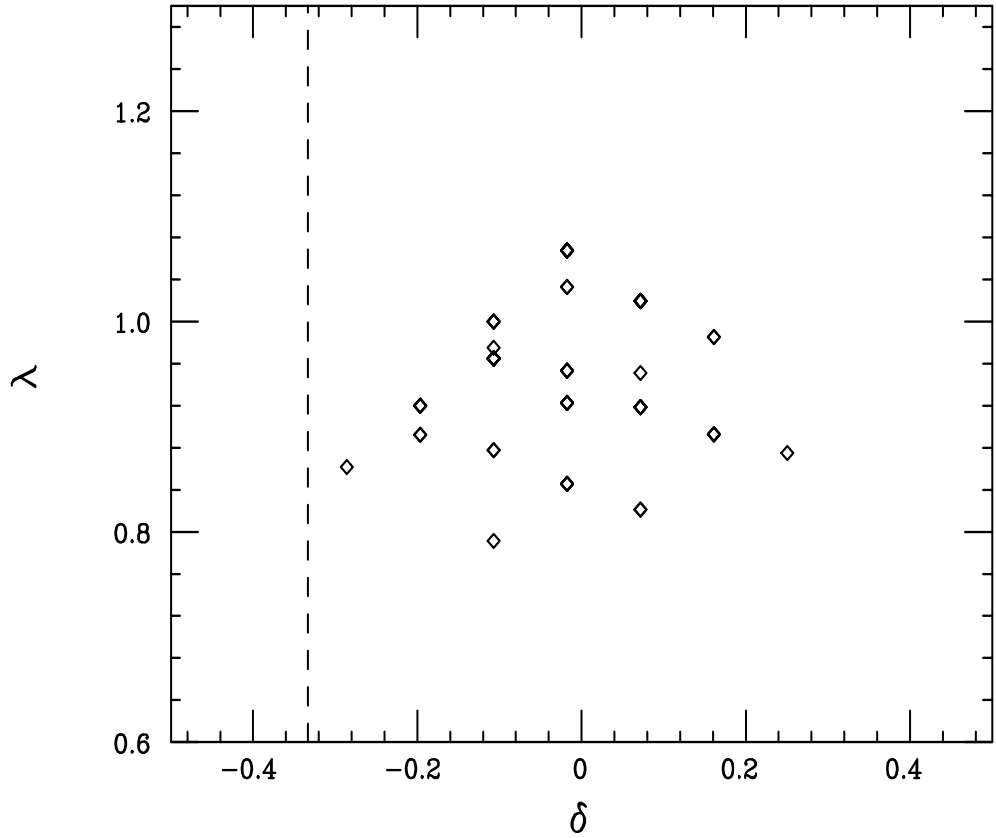,height=14cm,width=16cm,angle=-90}}
\vspace*{-1cm}
\caption{ Calculated values of the parameters 
$\delta$ and $\lambda$ for the 68=4+64 $E_6$ 
models from cases ($i$) and ($ii$) discussed in the text when 
$\eta$-type couplings are assumed. The 
vertical dashed line corresponds to exact leptophobia. Almost all points are 
multiply occupied.}
\label{dis}
\end{figure}
\vspace*{0.4mm}

As models with $\eta$-type couplings are the only potential candidates for 
leptophobia, it is interesting to know the explicit $\delta-\lambda$ 
correlation in this case. We display in Figure~\ref{dis} all of 
the 20 distinct solutions for $\delta$ and $\lambda$ assuming this value of 
$\theta$ for models of either case ($i$) or ($ii$). To access just how 
leptophobic these models can be we calculated the $Z'$ search reach in each 
case for both the Tevatron Run II and LHC following the procedure used to 
obtain Fig.~\ref{fig1}. At the Tevatron, except for the most leptophobic 
case, the search reaches lie in the range $724-932(970-1150)$ GeV for an 
integrated luminosity of 2(30)~$fb^{-1}$ and generally conforms to the usual 
expectations. In the most leptophobic case, (4,4,1), these values drop to 
only 524(778) GeV, which is not great but far from nonexistent. At the LHC with 
a luminosity of 100~$fb^{-1}$, the mass reaches for all but the most 
leptophobic case lie in the range $3305-4415$ GeV; this then drops to only 
2730 GeV in the (4,4,1) case. Again this limit is poor relative to the others 
but it is quite substantial. Thus although the (4,4,1) model is as close to 
leptophobia as possible, the $Z'$'s leptonic couplings remain large enough for 
this particle to be observed in Drell-Yan collisions but with a somewhat 
reduced reach.

\begin{figure}[htbp]
\centerline{
\psfig{figure=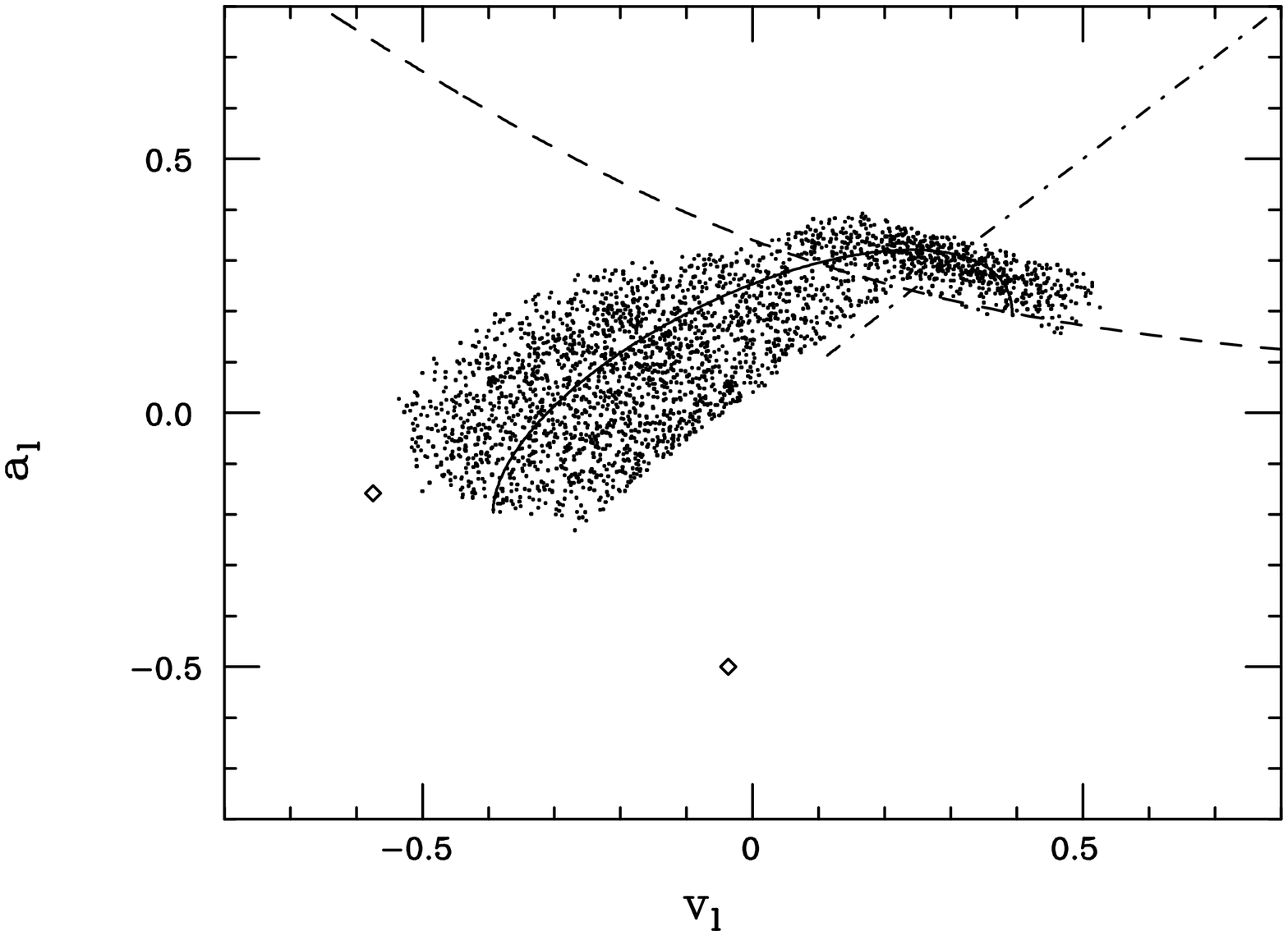,height=10.5cm,width=13cm,angle=-90}}
\vspace*{-5mm}
\centerline{
\psfig{figure=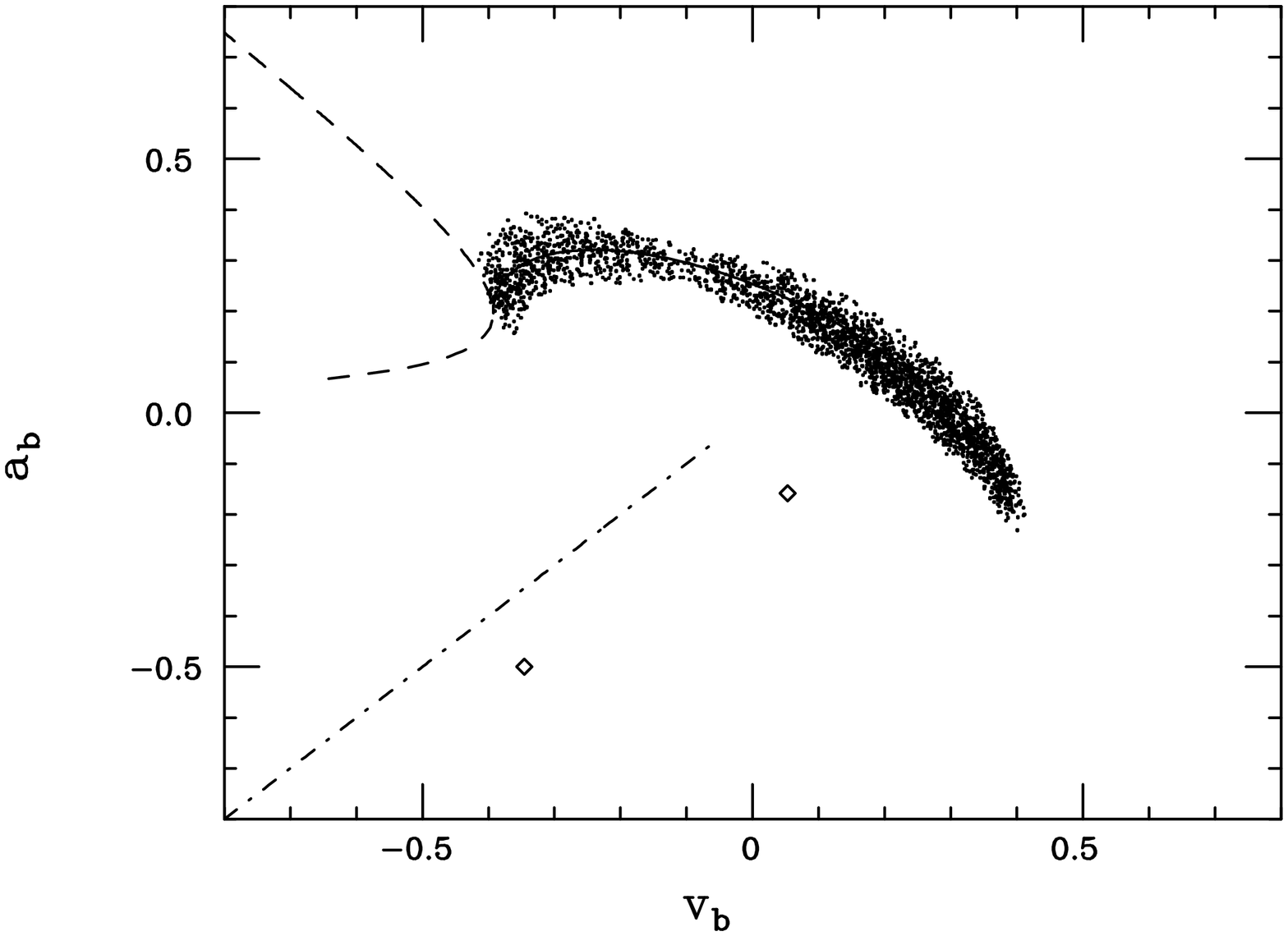,height=10.5cm,width=13cm,angle=-90}}
\vspace*{-0.9cm}
\caption{Calculated values for the vector and axial vector 
couplings for leptons(top) and 
$b$-quarks(bottom) arising from the 68 models $E_6$ models with kinetic 
mixing discussed in text in comparison to other $Z'$ models as in Fig.2. 
Complete $\theta-\delta-\lambda$ constraints and correlations are included.}
\label{fig5}
\end{figure}

Since the allowed ranges for both $\delta$ and $\lambda$ are reasonably 
restricted for these 68 models we would expect that the possibility of 
confusing a $E_6$ $Z'$ with that of a different model would be at least 
somewhat reduced. Figure~\ref{fig5} shows the regions of coupling parameter 
space allowed by the $\theta$-dependent $\delta$ and $\lambda$ constraints 
obtained above. The regions are seen to be somewhat smaller than those shown 
in the more pessimistic Fig.2 where $\lambda$ was set to unity and 
$\delta$ was free to vary over 
the range $-1/2 \leq \delta \leq 1/2$. There is certainly a significantly 
smaller overlap between the $E_6$ model predictions and those of other 
$Z'$ models making in likely that these classes of models would be 
distinguishable given sufficiently precise data and combining the results 
obtained for different flavor fermions. 

\vspace*{-0.5cm}
\nn
\vspace*{0.4mm}

\subsection{$SO(10)$}

In some sense the $SO(10)$ case is easier to deal with than is $E_6$ since 
here the parameter $\theta=-90^o$ is completely fixed. On the otherhand, the 
number of split multiplets that we can add at low energies is much larger 
thereby increasing the number of subcases to be examined. 
The reason for this is that, unlike $E_6$, the low energy content need only 
consist of the three {\bf 16}'s of $SO(10)$, plus the `Higgs' 
fields $H_1/H_1^c$ 
for the anomaly cancelation constraint to be satisfied. This means that, as 
in the MSSM, we may add ($i$) up to four 
${\bf 5}+\overline{\mbox{\bf 5}}$ ersatz pairs or ($ii$) 
one ${\bf 10}+\overline{\mbox{\bf 10}}$ either with or without an extra 
${\bf 5}+\overline{\mbox{\bf 5}}$ to this low energy content without a 
loss of perturbative unification. Of course in none of these cases will 
leptophobia be achieved but we will be able to constrain the 
range of allowed 
values for both $\delta$ and $\lambda$. Given this potentially large split 
multiplet field content at low energies it will be no surprise to find that 
these ranges are {\it significantly} larger than what was obtained above in 
the more constrained case associated with $E_6$.
For $n_5~{\bf 5}+\overline{\mbox{\bf 5}}$'s and 
$n_{10}~{\bf 10}+\overline{\mbox{\bf 10}}$'s, we already know from the MSSM 
that $\beta_L=1+n_5+3n_{10}$, $\beta_s=-3+n_5+3n_{10}$ and 
$B_{YY}=\beta_Y=33/5+n_5+3n_{10}$ with $n_5\geq 1$ and $n_{10}\geq 0$. 
Similarly, we also know the contribution of the three {\bf 16}'s to 
$B_{Q'Q'}=6$. To be more specific we need to examine the two individual cases 
independently. 

\begin{figure}[htbp]
\centerline{
\psfig{figure=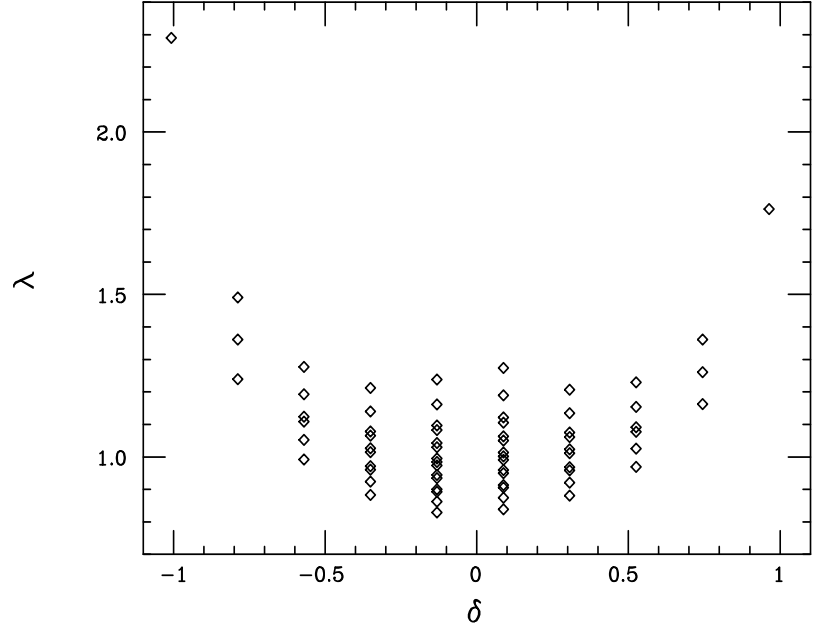,height=10.5cm,width=13cm,angle=-90}}
\vspace*{-5mm}
\centerline{
\psfig{figure=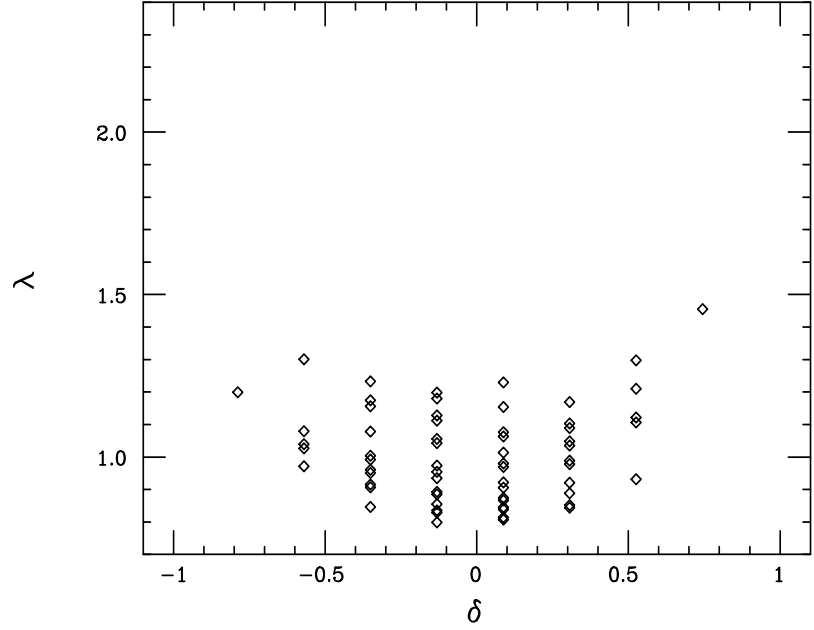,height=10.5cm,width=13cm,angle=-90}}
\vspace*{-0.9cm}
\caption{Distinct values of $\delta$ and $\lambda$ for the models associated 
with the two $SO(10)$ subcases ($i$)(top) and ($ii$)(bottom) discussed in 
the text.}
\label{fig6}
\end{figure}

In case ($i$) we are again dealing only with particles that lie in the 
${\bf 5}+\overline{\mbox{\bf 5}}$ as we did for $E_6$. The particle content 
can be thought of as $3\cdot{\bf 16}'s ~\oplus ~H_1/H_1^c ~\oplus 
~n[H_i/H_i^c+D_i/D_i^c]$ with $0\leq n\leq 4$.  Looking back at the $E_6$  
case we see that there are only two possible $\chi$ quantum number 
assignments for these fields: (1)~${\bf 5(2)}+\overline{\mbox{\bf 5(-2)}}$ and 
(2)~${\bf 5(-3)}+\overline{\mbox{\bf 5(3)}}$. For fixed $n$, we may have 
$(n_H,n_D)$ fields of type (1) and $(n+1-n_H,n-n_D)$ of type (2) with 
$0\leq n_H \leq n+1$ and $0\leq n_D\leq n$. Freely varying $n_{H,D}$ within 
their allowed ranges there are 2(6,12,20,30) subcases for $n=0$(1,2,3,4), for 
a total of 70. Here we find 
\begin{equation}
B_{YQ'}(i)= {6\over {5}}\sqrt {3\over {5}} [1+{5\over {3}}(n_D-n_H)]\,,
\end{equation}
and 
\begin{equation}
\Delta B_{Q'Q'}(i)=0.4n_H+0.6n_D+0.9(n_5+1-n_H)+1.35(n_5-n_D)\,.
\end{equation}
From these considerations we can immediately calculate the 
values of $\delta$ and $\lambda$ which depend upon $n,n_H$ and $n_D$; these 
are shown in the top part of Fig.~\ref{fig6}. Here we see that the results fill 
in a large crescent shaped region which extends to rather large values of 
both $\delta$ and $\lambda-1$ in comparison to the $E_6$ case as we 
anticipated from the large split multiplet content.

\vspace*{-0.5cm}
\nn
\vspace*{0.4mm}

In case ($ii$) we have a single ${\bf 10}+\overline{\mbox{\bf 10}}$ which may 
or may not be accompanied by an additional ${\bf 5}+\overline{\mbox{\bf 5}}$ 
so that $n_{10}=1$ and $0\leq n_5 \leq 1$. The possible $\chi$ quantum 
numbers of the ${\bf 5}+\overline{\mbox{\bf 5}}$ are given above and there are 
also two possibilities for the ${\bf 10}+\overline{\mbox{\bf 10}}$, \ie, 
${\bf 10(-1)}+\overline{\mbox{\bf 10(1)}}$ or 
${\bf 10(4)}+\overline{\mbox{\bf 10(-4)}}$. 
The particle content can be thought of symbolically as  
$3\cdot{\bf 16}'s ~\oplus ~H_1/H_1^c ~\oplus ~n_5[H_i/H_i^c+D_i/D_i^c]
~\oplus[Q/Q^c,E/E^c,U/U^c]$, with $0\leq n_5 \leq 1$ and, as in case ($i$), 
$0\leq n_H \leq n_5+1$ and $0\leq n_D \leq n_5$. If $n_Q(n_U,n_E)$ fields 
come from ${\bf 10(-1)}+\overline{\mbox{\bf 10(1)}}$ then $1-n_Q(1-n_U,1-n_E)$ 
come from ${\bf 10(4)}+\overline{\mbox{\bf 10(-4)}}$ since there is only one 
possible ${\bf 10}+\overline{\mbox{\bf 10}}$ allowed. Clearly 
$0\leq n_{Q,E,U} \leq 1$ independently of one another thus leading to a total 
of 64 subcases. We find for case ($ii$) the values 
\begin{equation}
B_{YQ'}(ii)= {6\over {5}}\sqrt {3\over {5}} [1+{5\over {3}}(n_D+2n_U-n_Q-
n_E-n_H)]\,,
\end{equation}
and 
\begin{equation}
\Delta B_{Q'Q'}(ii)=\Delta B_{Q'Q'}(i)+8-{3\over {4}}[n_E+3n_U+6n_Q]\,.
\end{equation}
From which we can immediately calculate the values of $\delta$ and 
$\lambda$; these are shown in the bottom part of Fig.~\ref{fig6}.  As in case 
($i$), the spread of $\delta$ and $\lambda$ values obtained for case ($ii$) 
remains significantly larger than in $E_6$ but less so than case ($i$). 

Due to the wide spread in the values of the $SO(10)$ $Z'$ couplings in the 
presence of KM, we may wonder if the hadron collider search reaches are 
drastically altered. At the Tevatron, we find that the search reaches lie 
in the range $824-938(1040-1208)$ GeV for an integrated luminosity of 
2(30)~$fb^{-1}$ and qualitatively conform to the usual model $\chi$ 
expectations, \eg, 864 GeV for 2~$fb^{-1}$, in the absence of KM. At the LHC 
with a luminosity of 100~$fb^{-1}$, the mass reaches lie in the range 
$4100-5315$ GeV again bracketing the non-KM expectation.

\vspace*{-0.5cm}
\nn
\begin{figure}[htbp]
\centerline{
\psfig{figure=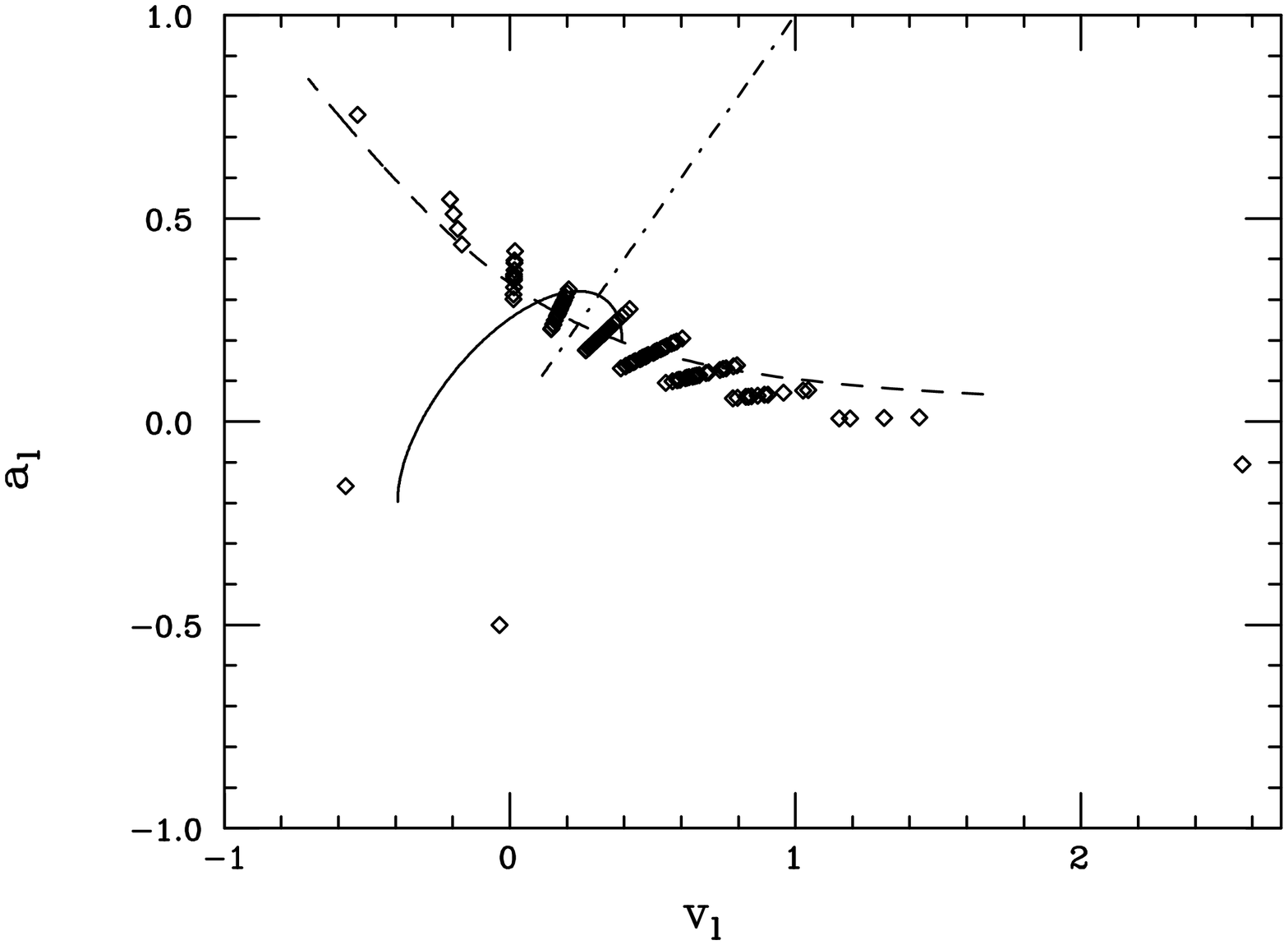,height=10.5cm,width=13cm,angle=-90}}
\vspace*{-5mm}
\centerline{
\psfig{figure=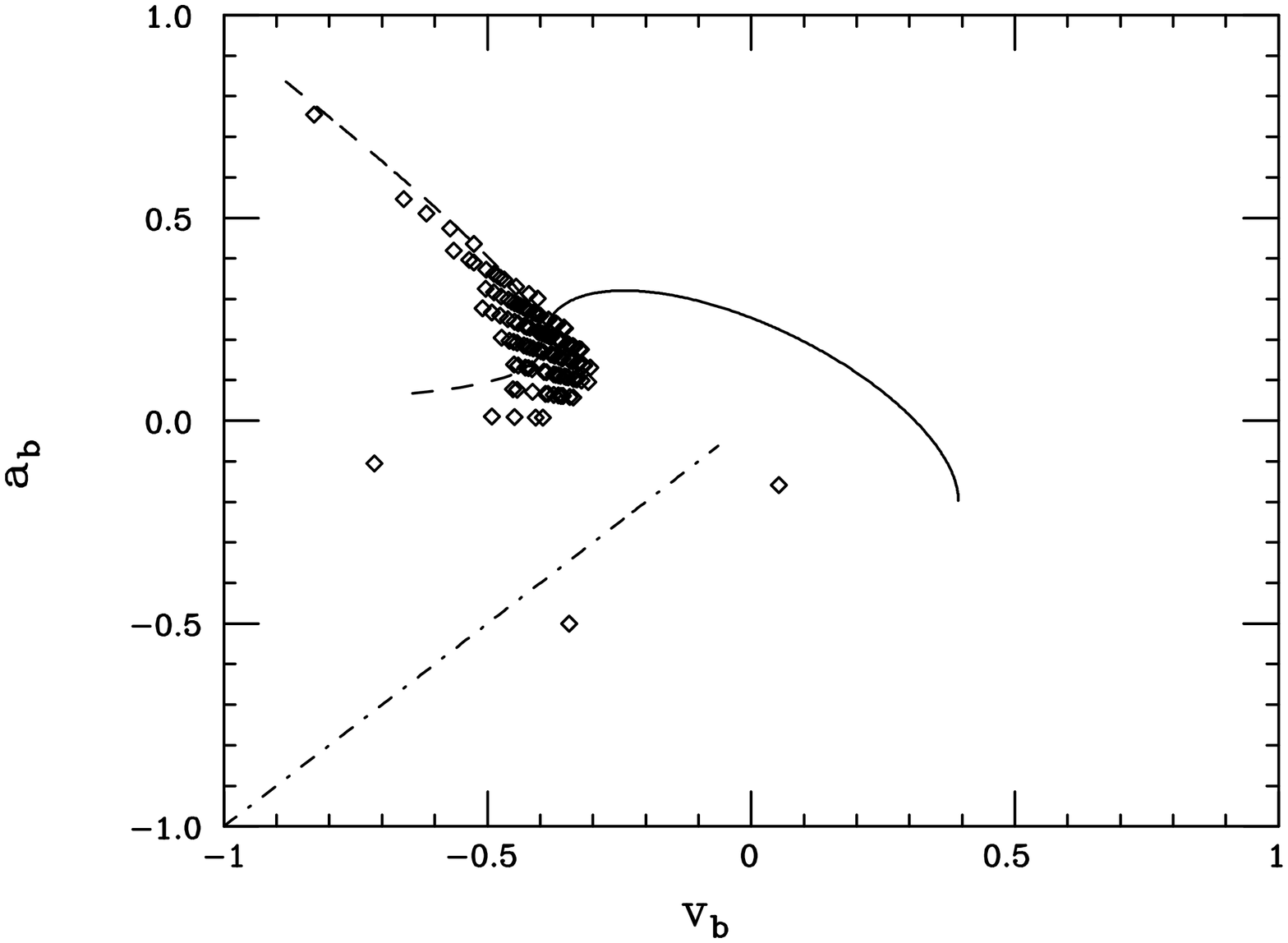,height=10.5cm,width=13cm,angle=-90}}
\vspace*{-0.9cm}
\caption{Predicted values of the vector and axial vector couplings in the 
134=64+70 $SO(10)$ cases discussed in the text compared with the predictions of 
other models as in Fig.2. Note the expanded scale in the present plots.}
\label{fig7}
\end{figure}
\vspace*{0.4mm}

What about the $Z'$ couplings themselves? These are completely specified by 
the values of $\delta$ and $\lambda$ and are shown in 
Figure~\ref{fig7} for all of the 134 subcases. Notice that they span a large 
range but tend to cluster near, but not necessarily on top of, those of the 
LRM. We recall from our earlier discussion than in the absence of KM the 
model $\chi$ couplings are exactly the same as those of the LRM with 
$\kappa^2=(g_R/g_L)^2={5\over {3}} {x_w\over {(1-x_w)}}$. It is apparent from 
the figure that once KM is turned on there is no obvious relationship between 
the two sets of couplings, though they do seem to track one another. A short 
analysis, however, shows that there does exist a value of $\kappa$ in the LRM 
for which the {\it ratios} of couplings are the same as in the $\chi$ model 
with KM for any value of $\delta$. This means that for this value of $\kappa$ 
the couplings of the LRM and $\chi$ model with KM are identical apart from an 
overall normalization. This is easily proved by considering both the general 
form of the LRM couplings and remembering that the value of $Q_\chi$ can be 
written as a linear combination of $T_{3R}$ and $Y$, where $T_{3R}$ is the 
third component or the right-handed weak isospin. Specifically we find that 
correspondence in the couplings between the two models occurs when 
\begin{equation}
\kappa^2={5x_w\over {3(1-x_w)}}[1-\sqrt 6 \delta /3]^{-1}\,,
\end{equation}
and we see the conventional well-known result is recovered when $\delta \to 0$. 
Except for when $\delta$ is large and negative, $\simeq -1$, this equation 
has a solution in the physical region of the LRM, \ie, 
$\kappa^2 > {x_w\over {(1-x_w)}}$. This 
result has very important implications to issues involving around $Z'$ 
coupling determinations at colliders.

As is well known, most techniques 
aimed at identifying a new $Z'$ at hadron colliders{\cite {steve,snow}} 
actually employ observables which only determine various ratios of fermionic 
couplings. The extraction of coupling information from other observables, 
such as the $Z'$ total width, are not only subject to larger systematic errors 
but depend on assumptions about how the $Z'$ can decay. As we have just 
seen the ratios of $SO(10)$-inspired $\chi$ model couplings in the presence 
of KM can be easily mimicked by those of the LRM with a suitably chosen value 
of the $\kappa$ parameter. Thus the $Z'$'s of these two models could be easily 
confused.

\begin{figure}[htbp]
\centerline{
\psfig{figure=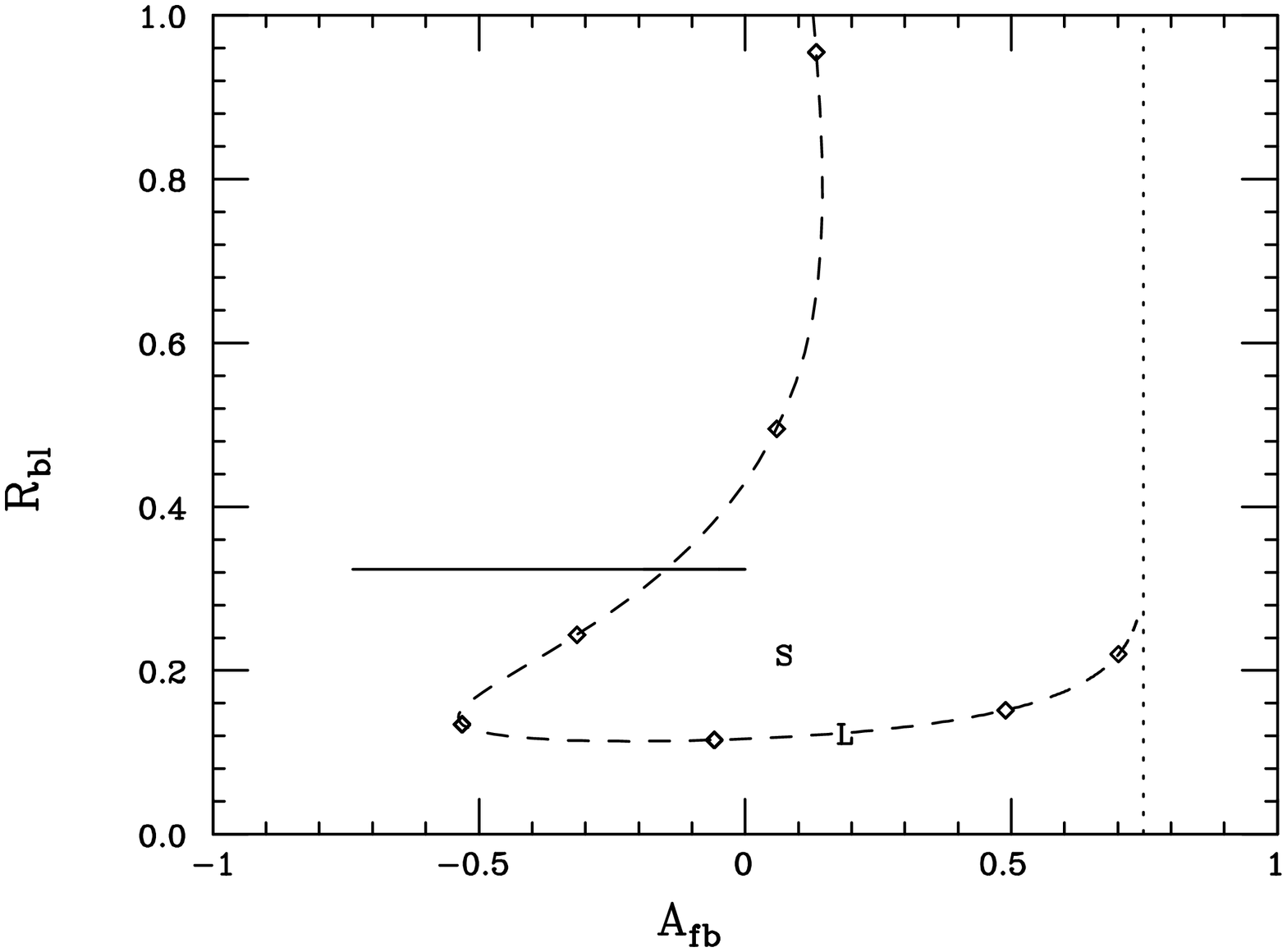,height=10.5cm,width=13cm,angle=-90}}
\vspace*{-5mm}
\centerline{
\psfig{figure=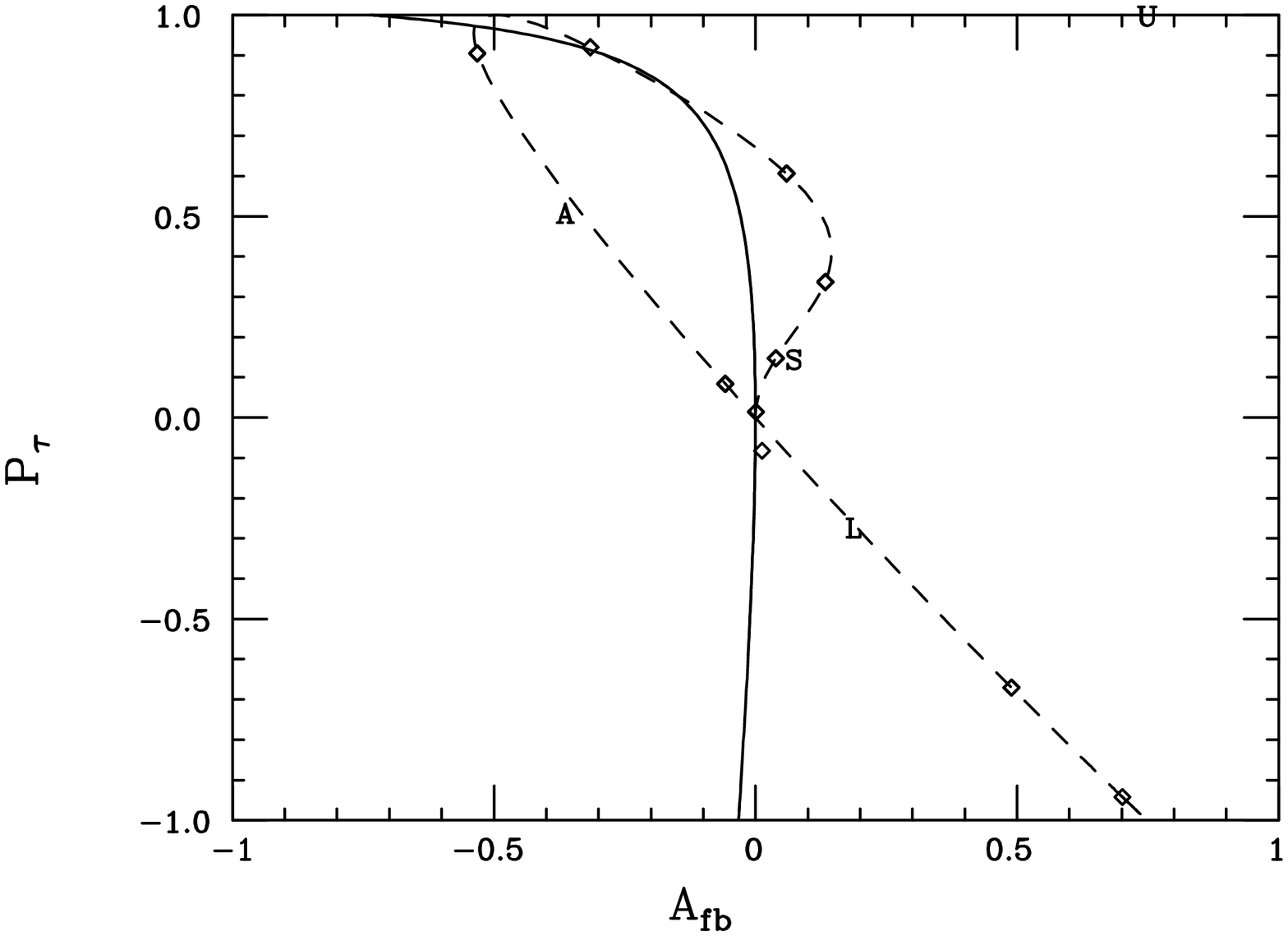,height=10.5cm,width=13cm,angle=-90}}
\vspace*{-0.9cm}
\caption{Correlations in the values of observables used to extract $Z'$ 
coupling information at hadron colliders as discussed in the text. 
The diamonds are the predictions of the $SO(10)$-inspired 
$\chi$ model with KM. The solid(dashed,dotted) curves correspond to the 
$E_6$ model without KM, the LRM and the Un-unified Model, respectively. 
The letters `A,L,U,S' label the predictions for Ma's model, the LRM with 
$\kappa=1$, the Un-unified Model and a heavy SM $Z'$, respectively.}
\label{fig8}
\end{figure}

As an example of this, let us consider the production of a $\simeq 700$ GeV 
$Z'$ at the Tevatron during Run II. After a few $fb^{-1}$ of luminosity are 
available several 100's of events in the dilepton channel will have been 
collected. Give the limited statistics, only a few of the variously proposed 
observables can be used to examine the $Z'$ coupling. In addition to the 
charged lepton forward-backward asymmetry, $A_{fb}$, one might measure the 
relative cross section in $b\bar b$ final states, $R_{bl}$, as suggested 
by{\cite {pkm}}, or the polarization of one of the $\tau$'s, $P_\tau$, in 
$Z'\to \tau^+\tau^-$, as suggested by{\cite {cahn}}. Fig.~\ref{fig8} shows 
the correlations amongst these observables and, in particular, compares the 
LRM predictions with those of the $\chi$ model with kinetic mixing. We see 
immediately that the two models would be quite easily confused. Of course, 
in the LRM case a $W'$ also exists with a mass somewhat less than the $Z'$; 
finding the $W'$ may be the only way to distinguish these two cases. (In some 
cases, finding the $W'$ may also be difficult{\cite {snow}}.)

At lepton colliders operating below the $Z'$ production threshold, 
measurements made at a single $\sqrt s$ are insensitive to the overall 
normalization of the $Z'$ couplings. Their apparent values can be easily 
adjusted by 
a simple rescaling of the $Z'$ mass. This weakness can be overcome at lepton 
colliders, however, by combining measurements taken at several distinct 
values of $\sqrt s${\cite {snow,me2}}. Thus, in principle, lepton colliders 
can be used to distinguish the LRM and $\chi$ model with KM cases. Of course, 
if such a machine can operate on the $Z'$ pole and the coupling normalization 
determined, there will be no ambiguities in $Z'$ model identification. 

\vspace*{-0.5cm}
\nn
\vspace*{0.4mm}

\section{Discussion and Summary}

In this paper we have performed a detailed examination of the magnitude and 
influence of gauge kinetic mixing on couplings of the $Z'$'s which originate 
from either $E_6$ or $SO(10)$. These mixing effects were shown to be 
completely described by the values of the two parameters $\delta$ and 
$\lambda$ which can be obtained via a renormalization group analysis. After 
introducing several model building assumptions we numerically analyzed the 
68 $E_6$ and 134 $SO(10)$ models to which these assumptions naturally led. The 
values of both $\delta$ and $\lambda$ were calculated for both sets of models, 
in particular, as functions of the mixing angle $\theta$ in the $E_6$ cases. 

For the $E_6$ models, since the number of additional low energy matter 
representations inducing kinetic mixing was constrained to be rather limited 
due to our model building assumptions, 
the allowed ranges of both parameters was shown to be quite restricted. 
Futhermore, we demonstrated that exact leptophobia, which occurs when 
$\delta=-1/3$ for the conventional $E_6$ particle embedding in model $\eta$, 
is impossible to achieve in any of these models. This result was shown to 
be independent of how the fermions and additional vector-like matter fields 
necessary to induce kinetic mixing are embedded in GUT representations. In 
the case which was closest to being leptophobic, we determined that the 
leptonic couplings of the $Z'$ were sufficiently large to render it visible 
in Drell-Yan collisions at both the Tevatron and the 
LHC. Of course in comparison to models where kinetic mixing is absent the 
reach for such a $Z'$ was found to be significantly reduced by $\simeq 40\%$. 
Furthermore, in the general $E_6$ case, we showed that the couplings of the 
$Z'$ remain sufficiently distinct from those of other models, such as the 
Left-Right Model, that they could be easily identified once sufficient 
statistics becomes available at future colliders. We demonstrated that this 
result would not hold if the magnitude of the kinetic mixing contributions to 
the $Z'$ couplings were left unrestricted.

For the $SO(10)$-inspired $\chi$ models, the potential effects of kinetic 
mixing were shown to be more pronounced (though leptophobia can never arise in 
these scenarios). This is due to the much larger range of split multiplets that 
may be introduced in this case while still satisfying our model building 
assumptions. In many cases kinetic mixing was shown to lead to values of 
$\lambda$ significantly greater than unity which resulted in increased 
discovery reaches for these $Z'$ at both the Tevatron and LHC. Qualitatively, 
the significantly 
expanded range of allowed $\chi$ couplings were found to track those of 
the LRM. In particular, we demonstrated that for all allowed values of $\delta$ 
there exists a corresponding 
value of the LRM parameter $\kappa$ for which the couplings 
in the two theories are identical apart from an overall normalization. This 
was shown to have a serious impact on $Z'$ model discrimination at hadron 
colliders as well as at lepton colliders unless data taken at multiple 
$\sqrt s$ values is available foe analysis.

The influence of gauge kinetic mixing leads to an enrichment in the 
phenomenology of new gauge bosons. Hopefully such particles will be found 
at future colliders.

\noindent{\Large\bf Acknowledgements}

The author would like to thank J.L. Hewett, J. Wells and D. Pierce for 
discussions related to this work.

\newpage

%
\def\MPL #1 #2 #3 {Mod. Phys. Lett. {\bf#1},\ #2 (#3)}
\def\NPB #1 #2 #3 {Nucl. Phys. {\bf#1},\ #2 (#3)}
\def\PLB #1 #2 #3 {Phys. Lett. {\bf#1},\ #2 (#3)}
\def\PR #1 #2 #3 {Phys. Rep. {\bf#1},\ #2 (#3)}
\def\PRD #1 #2 #3 {Phys. Rev. {\bf#1},\ #2 (#3)}
\def\PRL #1 #2 #3 {Phys. Rev. Lett. {\bf#1},\ #2 (#3)}
\def\RMP #1 #2 #3 {Rev. Mod. Phys. {\bf#1},\ #2 (#3)}
\def\ZPC #1 #2 #3 {Z. Phys. {\bf#1},\ #2 (#3)}
\def\IJMP #1 #2 #3 {Int. J. Mod. Phys. {\bf#1},\ #2 (#3)}

\end{document}